\documentclass[sigconf,natbib=false,screen,authorversion]{acmart} 

\AtBeginDocument{%
  }

\copyrightyear{2024}
\acmYear{2024}
\setcopyright{acmlicensed}
\acmConference[CCS '24]{Proceedings of the 2024 ACM SIGSAC Conference on Computer and Communications Security}{October 14--18, 2024}{Salt Lake City, UT, USA}
\acmBooktitle{Proceedings of the 2024 ACM SIGSAC Conference on Computer and Communications Security (CCS '24), October 14--18, 2024, Salt Lake City, UT, USA}
\acmDOI{10.1145/3658644.3670368}
\acmISBN{979-8-4007-0636-3/24/10}

\RequirePackage[
  datamodel=acmdatamodel,
  style=acmnumeric,
  ]{biblatex}

\usepackage{hyperref}
\usepackage{balance}
\usepackage{subcaption}
\usepackage[boxed,vlined,linesnumbered]{algorithm2e}
\usepackage{nicefrac}

\newcommand{\qm}[1]{``#1''}

\newcommand{\bh}{\texttt{browsing\_history}}
\newcommand{\context}{\texttt{context}}
\newcommand{\timestamp}{\texttt{timestamp}}
\newcommand{\origin}{\texttt{origin}}
\newcommand{\cookie}{\texttt{cookie}}
\newcommand{\cookies}{\texttt{cookies}}
\newcommand{\uid}{\texttt{uid}}
\newcommand{\taxonomy}{\texttt{taxonomy}}
\newcommand{\classifier}{\texttt{classifier}}

\newtheorem{theorem}{Theorem}
\newtheorem{definition}[theorem]{Definition}

\newtheorem{lemma}[theorem]{Lemma}
\newtheorem{corollary}[theorem]{Corollary}

\newtheorem{remark}[theorem]{Remark}

\begin{document}

\title{The Privacy-Utility Trade-off in the Topics API}

\author{Mário S.\ Alvim}
\email{msalvim@dcc.ufmg.br}
\orcid{0000-0002-4196-7467}
\affiliation{%
  \institution{Universidade Federal de Minas Gerais}
  \city{Belo Horizonte}
  \state{MG}
  \country{Brazil}
}

\author{Natasha Fernandes}
\email{natasha.fernandes@mq.edu.au}
\orcid{0000-0002-9212-7839}
\author{Annabelle McIver}
\email{annabelle.mciver@mq.edu.au}
\orcid{0000-0002-2405-9838}
\affiliation{%
  \institution{Macquarie University}
  \city{Sydney}
  \state{NSW}
  \country{Australia}
}

\author{Gabriel H.\ Nunes}
\email{ghn@nunesgh.com}
\orcid{0000-0002-7823-3061}
\affiliation{%
  \institution{Macquarie University}
  \city{Sydney}
  \state{NSW}
  \country{Australia}
}
\affiliation{%
  \institution{Universidade Federal de Minas Gerais}
  \city{Belo Horizonte}
  \state{MG}
  \country{Brazil}
}

\renewcommand{\shortauthors}{Mário S.\ Alvim, Natasha Fernandes, Annabelle McIver, \& Gabriel H.\ Nunes}

\begin{abstract}
    The ongoing deprecation of third-party cookies by web browser vendors has sparked the proposal of alternative methods to support more privacy-preserving personalized advertising on web browsers and applications.
    The Topics API is being proposed by Google to provide third-parties with \qm{coarse-grained advertising topics that the page visitor might currently be interested in}.
    In this paper, we analyze the {re-iden\-ti\-fi\-ca\-tion} risks for individual Internet users and the utility provided to advertising companies by the Topics API, i.e. learning the most popular topics and distinguishing between real and random topics.
    We provide theoretical results dependent only on the API parameters that can be readily applied to evaluate the privacy and utility implications of future API updates, including novel general upper-bounds that account for adversaries with access to unknown, arbitrary side information, the value of the differential privacy parameter $\epsilon$, and experimental results on real-world data that validate our theoretical model.
\end{abstract}




\maketitle

\section{Introduction}
\label{sec:introduction}

The ongoing deprecation of third-party cookies by web browser vendors has sparked the proposal of alternative methods to support more privacy-preserving personalized advertising on web browsers and applications.
For instance, Google's Privacy Sandbox initiative for the web \cite{Schuh2019} and for Android \cite{Chavez2022} initially included FLoC \cite{Google2019} and the Protected Audience API \cite{Google2020} (once FLEDGE \cite{Southey2023}).
The Topics API \cite{Goel2022,Google2022}, which we analyze here, was later introduced to replace FLoC.
Other initiatives include Microsoft's PARAKEET \cite{Microsoft2021} and MaCAW \cite{Microsoft2021a}, and IPA \cite{Mozilla2022} by both Meta and Mozilla \cite{Thomson2022}.

Even though Google continues to deploy the Topics API on both Chrome browser and Android, the specification was opposed by other major web browser vendors, e.g. Apple \cite{WebKitOpenSourceProject2022} and Mozilla \cite{Mozilla2023,Thomson2023}, and was not accepted by the \emph{W3C Technical Architecture Group} (W3CTAG) \cite{W3CTechnicalArchitectureGroup2023}.
In particular, we highlight:
(i) W3CTAG's concerns on the use of side information in addition to data provided by the API and the lack of a limit on the size of the taxonomy of topics, and
(ii) Mozilla's concerns on the privacy impact for individuals on worst-case scenarios, the possible inefficacy of the differential privacy aspects of the API, and the actual utility of the API for \emph{interest-based advertising} (IBA) companies.



\paragraph{Objectives}
The Topics API is expected to \qm{support IBA without relying on cross-site tracking} \cite{Carey2023} by making it harder to link topics observed on distinct websites to the same individual, if compared to the direct linkage enabled by third-party cookies.
In fact, results published by Google estimate the probability of a correct re-identification of a random individual would be below $3\%$ \cite{Carey2023}.

Our goal is to formally analyze the Topics API by developing sound yet easily explainable models for both the API and third-party cookies, which we use as a baseline.
We are interested in both the privacy (for Internet users) and utility (for IBA companies) implications of using the API, how they relate, and their trade-off.

\paragraph{Contributions}
In summary, our main contributions are:
\begin{itemize}
    \item We propose a novel model for the Topics API based on the theory of \emph{Quantitative Information Flow} (QIF).
    \item We provide novel theoretical privacy results for individuals, e.g. privacy vulnerability and upper-bounds for average- and worst-case scenarios, evaluation of the differential privacy parameter $\epsilon$, and use of unknown, arbitrary correlations as side information, all dependent only on the API parameters.
    \item We provide novel theoretical utility results for IBA, i.e. learning the most popular topics and distinguishing between real and random topics.
    \item We provide novel datasets for analyses of Internet users' browsing histories and topics of interest.
    \item We provide experimental results on real-world data that validate the predictions of our theoretical models.
\end{itemize}

Further, our theoretical results can be readily applied to evaluate the privacy and utility implications of future updates to the API.

\newpage

\paragraph{Plan of the paper.}
A background on third-party cookies, the Topics API, and QIF, is in Sec.~\ref{sec:technical}.
Our formal models for third-party cookies and the Topics API are in Sec.~\ref{sec:models} and the theoretical results for their privacy and utility are in Sec.~\ref{sec:theoretical}.
Experimental evaluations that corroborate our theoretical results are in Sec.~\ref{sec:experimental} and discussion is in Sec.~\ref{sec:discussion}.
Related work is in Sec.~\ref{sec:related-work} and conclusion is in Sec.~\ref{sec:conclusion}.
Proofs are in App.~\ref{app:proofs}, detailed steps on data treatment are in App.~\ref{app:datasets}, and how to extend our Topics API model to account for more than one epoch is in App.~\ref{app:model-longitudinal}.

\section{Technical Background}
\label{sec:technical}

\begin{figure*}[ht]
    \centering
    \begin{subfigure}[b]{0.45\textwidth}
        \includegraphics[width=1.00\textwidth]{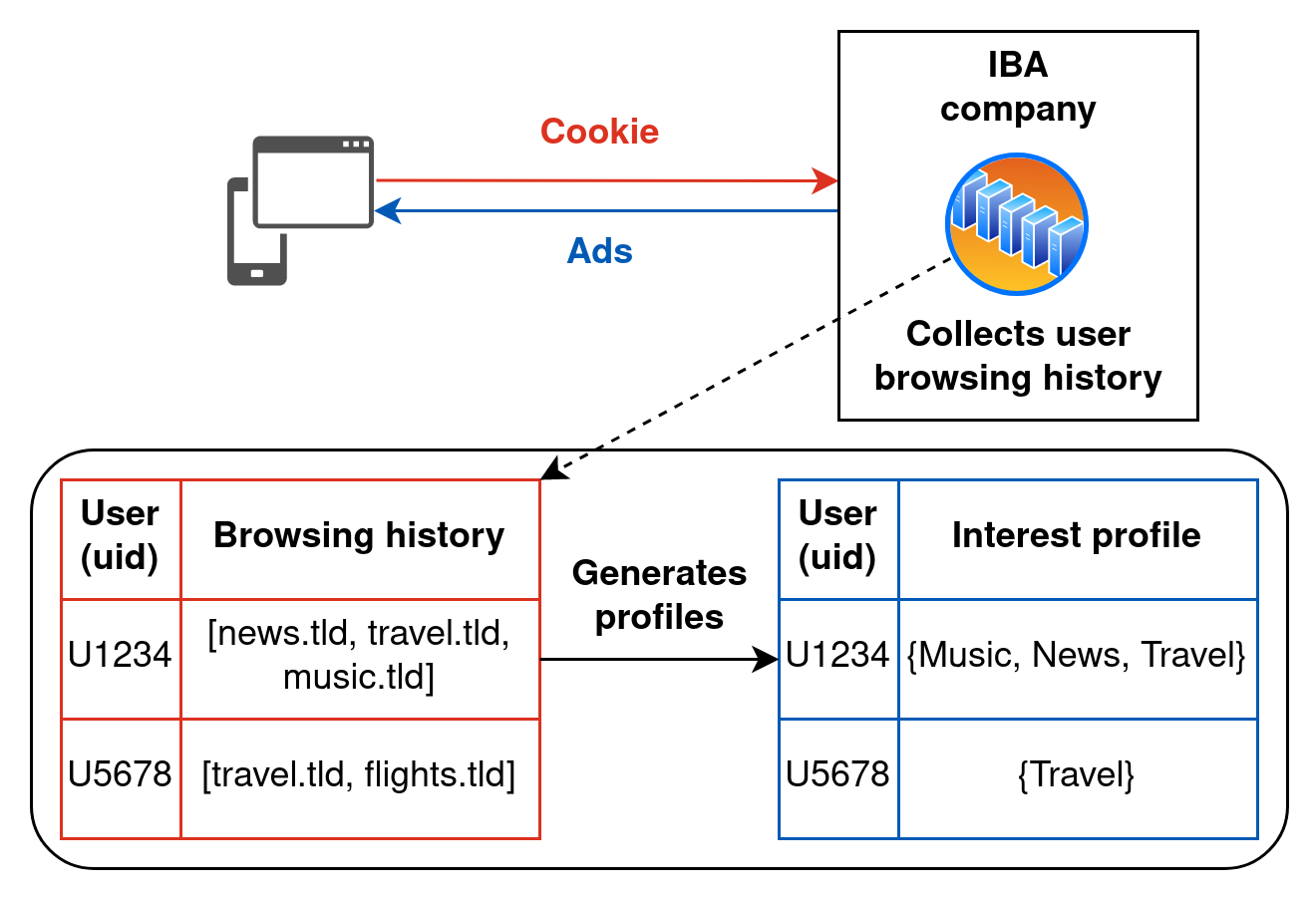}
        \caption{An IBA company using third-party cookies.}
        \label{fig:cookies}
    \end{subfigure}
    \hfil
    \begin{subfigure}[b]{0.45\textwidth}
        \includegraphics[width=1.00\textwidth]{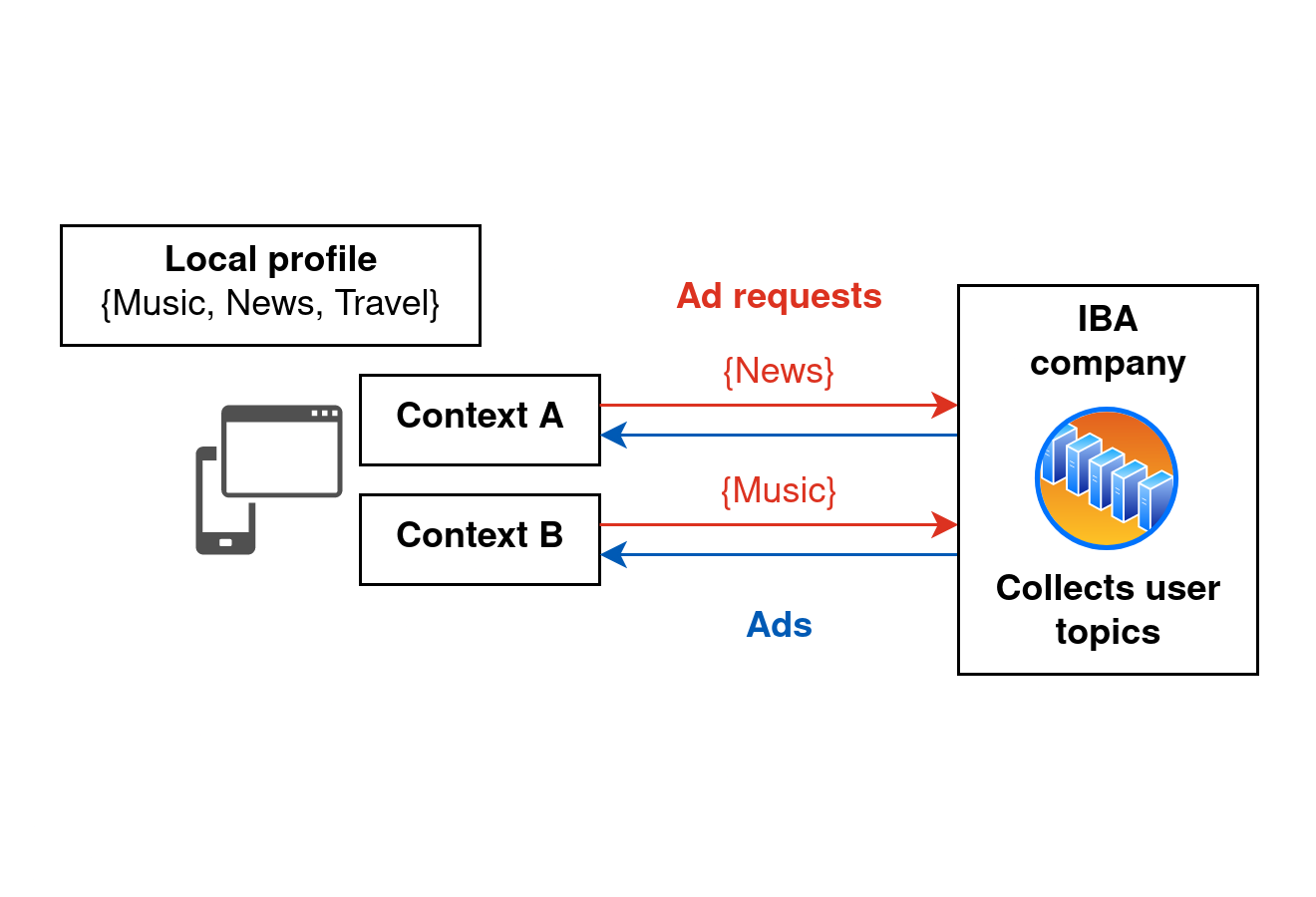}
        \caption{An IBA company using the Topics API.}
        \label{fig:topics}
    \end{subfigure}
    \caption{Information collected by an interest-based advertising (IBA) company.}
    \Description{An IBA company using third-party cookies collects detailed browsing histories for each user in order to generate interest profiles. The same company using the Topics API would collect only topics of interest for each user.}
    \label{fig:schematics}
\end{figure*}

We review third-party cookies in Sec.~\ref{sec:technical-cookies}, the Topics API in Sec.~\ref{sec:technical-topics}, and QIF in Sec.~\ref{sec:technical-qif}, so we can present our models in Sec.~\ref{sec:models}.

\subsection{Third-party Cookies}
\label{sec:technical-cookies}

\begin{algorithm}[ht]
    \DontPrintSemicolon
    \KwOut{\texttt{(}\cookie, \context, \timestamp\texttt{)} to \origin.}
    \BlankLine
    
    \SetKwFunction{ReportCookieAndContext}{ReportCookieAndContext}
    \SetKw{In}{in}
    
    \bh\ $\leftarrow$ (\context, \timestamp) \tcp*[r]{$\mathsf{C}_{\mathit{BH}}$}
    
    \uIf(\tcp*[f]{$\mathsf{C}_{U}$}){\origin\ \In \cookies}{
        \Return{\origin, \texttt{(}\cookie, \context, \timestamp\texttt{)}}\;
    }
    \Else{
        \cookies\ $\leftarrow$ \{\origin: \cookie\}\;
        \Return{\origin, \texttt{(}\cookie, \context, \timestamp\texttt{)}}\;
    }
    \caption{Browser's perspective of third-party cookies. All variables are global. $\mathsf{C}_{\mathit{BH}}$ and $\mathsf{C}_{U}$ refer to information-theoretical channels that model the respective sections of the algorithm and which are formally defined in Sec.~\ref{sec:models-cookies}. Here, \context\ is the web page being currently visited, \origin\ is any domain called from the current one (itself included), \bh\ is a list of visited web pages, and \cookies\ is a dictionary mapping domains to their \cookie.}
    \label{alg:cookies}
\end{algorithm}

Cookies were first formally specified by the \emph{Internet Engineering Task Force} (IETF) in 1997 as \qm{a way to create a stateful session with HTTP requests and responses} \cite{Montulli1997}.
From the start, it was known that cookies were vulnerable to privacy abuse through \qm{cookie sharing}, i.e. \emph{third-party cookies}, and web browser vendors were \qm{strongly encouraged} to \qm{prevent the sharing of session information between hosts that are in different domains} \cite{Montulli1997}, but more strict cookies policies were only introduced in the early 2010s \cite{Mayer2013}.

A \emph{cookie} is a piece of information stored by a web browser that consists of a tuple of the \emph{origin} (domain) that set that cookie and one or more pairs of keys and values.
For instance, a cookie may be used to store the content of a shopping cart on a e-commerce origin, or the login information on a social network or e-mail service.
But a cookie may also be set by any third-party origin called on a \emph{context} (visited web page), e.g. through advertisements or social widgets.
If a third-party sets a cookie with a unique identifier (\uid) on a browser, it becomes capable of tracking the browsing history of that individual on that browser whenever the third-party is called across the Internet, as in Fig.~\ref{fig:cookies}, which enables the creation of precise browsing profiles for individuals at scale \cite{Roesner2012,Englehardt2015,Lerner2016,Bashir2018}.
Browser vendors started to deprecate third-party cookies in 2019 \cite{Wilander2020,Wood2019}.

We represent the behavior of third-party cookies from a browser's perspective in Alg.~\ref{alg:cookies}, which runs whenever an Internet user visits a web page (\context).
For each visited context, zero or more third-party domains (\origin) may be called, which would trigger Alg.~\ref{alg:cookies} to send to that origin its (possibly just created) \cookie, i.e the \uid, the \context\ from which the call was made, and the \timestamp.

\subsection{Topics API}
\label{sec:technical-topics}

\begin{algorithm}[ht]
    \DontPrintSemicolon
    \SetKwFunction{GenerateTopics}{GenerateTopics}
    \SetKwFunction{ComputeTopTopics}{ComputeTopTopics}
    \SetKw{In}{in}
    
    \KwIn{$s$: size of the top-$s$ topics set.}
    \KwOut{$top\_s$: top-$s$ topics set for current epoch.}
    \BlankLine
    
    \For(\tcp*[f]{$\mathsf{C}_{G}$}){\context\ \In \bh}{
        $\mathit{topics} \leftarrow$ \{\context: \GenerateTopics{\context, \taxonomy, \classifier}\}\;
    }
    
    $top\_s \leftarrow$ \ComputeTopTopics{$\mathit{topics}$, $s$}\;
    \Return{$top\_s$}
    \caption{Topics API computation of top-$s$ topics from a browser's perspective. All variables are global, except those in \emph{italic}, which are local to the algorithm. $\mathsf{C}_{G}$ refers to an information-theoretical channel that models the whole algorithm and which is formally defined in Sec.~\ref{sec:models-topics}. Here, $\mathit{topics}$ is a dictionary mapping visited domains to their topics.}
    \label{alg:topics-derive}
\end{algorithm}

\begin{algorithm}[ht]
    \DontPrintSemicolon
    \SetKwBlock{With}{with probability}{end}
    \SetKwFunction{UniformRandomTopic}{UniformRandomTopic}
    
    \KwIn{$r$: probability of returning a random topic from the whole taxonomy; $top\_s$: topics set for current epoch.}
    \KwOut{$\mathit{topic}$: reported topic.}
    \BlankLine
    
    $\mathit{topic} \leftarrow$ \UniformRandomTopic{$top\_s$} \tcp*[r]{$\mathsf{C}_{\mathit{BN}}$}
    \With($r$\tcp*[f]{$\mathsf{C}_{\mathit{BN} \oplus_{r} \mathit{DP}}$}){
        $\mathit{topic} \leftarrow$ \UniformRandomTopic{\taxonomy}\;
    }
    \Return{$\mathit{topic}$}
    \caption{Topics API topic reporting from a browser's perspective. All variables are global, except those in \emph{italic}, which are local to the algorithm. $\mathsf{C}_{\mathit{BN}}$ and $\mathsf{C}_{\mathit{BN} \oplus_{r} \mathit{DP}}$ refer to information-theoretical channels that model the respective sections of the algorithm and which are formally defined in Sec.~\ref{sec:models-topics}.}
    \label{alg:topics-report}
\end{algorithm}

Considering the deprecation of third-party cookies, Google has proposed the Topics API to provide third-parties with \qm{coarse-grained advertising topics that the page visitor might currently be interested in} \cite{Google2022}.
This includes \emph{interest-based advertising} (IBA), which differs from contextual advertising by taking into account not only the context for deriving an individual's set of interests, but also other signals, e.g. interest profiles derived from browsing histories collected via third-party cookies.

The Topics API proposes the representation of an individual as a set of top interests (\emph{topics}) locally derived from their browsing history.
It also includes a pre-defined taxonomy of topics and a pre-trained classification model mapping contexts to topics.
Roughly, at the end of each \emph{epoch} (e.g. week), the individual's browser computes a fixed-size ($s$) set of top topics based on the browsing history and the topics assigned to each of the observed contexts by the classifier.
Once computed, this set would be available to third-parties for a fixed number of epochs and under certain restrictions, e.g. Fig.~\ref{fig:topics}.

For instance, a third-party would receive only one topic per individual, per epoch, and per context, chosen uniformly at random from that individual's set of top topics for that epoch.
Moreover, a third-party would not receive a topic unrelated to the contexts it has witnessed that individual visit on that epoch, and, with a $5\%$ chance, the received topic would be instead chosen uniformly at random from the whole taxonomy \cite{Carey2023,Google2022}.

We represent the behavior of the Topics API from a browser's perspective in Alg.~\ref{alg:topics-derive}, which runs once per epoch to derive the top-$s$ topics set for that epoch, and Alg.~\ref{alg:topics-report}, which runs whenever an Internet user visits a web page (\context) that calls the Topics API, assuming the caller satisfies the requirements to receive a $\mathit{topic}$.
For each run of Alg.~\ref{alg:topics-report}, a $\mathit{topic}$ is reported after being chosen uniformly at random from the top-$s$ topics set, which may be overridden with probability $r$ by a topic chosen uniformly at random from the whole taxonomy.
We assume the browsing history used in Alg.~\ref{alg:topics-derive} was already collected during each epoch as on line 1 of Alg.~\ref{alg:cookies}.

\subsection{Quantitative Information Flow}
\label{sec:technical-qif}

Quantitative Information Flow (\emph{QIF}) is an information- and decision-theoretic framework \cite{Alvim2020} that has already been successfully applied to privacy and security analyses, e.g. searchable encryption \cite{Jurado2019}, intersection and linkage attacks against $k$-anonymity \cite{Fernandes2018,Alvim2022}, privacy analysis of very large datasets \cite{Nunes2021,Alvim2022}, and differential privacy \cite{Alvim2015,Chatzikokolakis2019,Alvim2020a,Alvim2023}.
In this paper, we use QIF to formally model the information leakage caused by both the Topics API and third-party cookies.


QIF models information leakage for Bayesian adversaries.
It is distinguished by its use of the \emph{g}-vulnerability framework, which enables information leakage to be measured relative to different adversarial scenarios.
The result of a QIF analysis is an average- or worst-case estimate of how much an adversary is able to use the leaked information, e.g. for performing a subsequent linkage attack.
The framework is flexible in that we can analyze both privacy vulnerabilities and the impact of privacy defenses on utility.

\paragraph{Prior Vulnerability}
The first step in a QIF analysis is to determine the adversary's prior knowledge of the secret information that they would like to learn.
A \emph{secret} in this sense is modeled as a probability distribution $\pi$ of type $\mathbb{D} \mathcal{X}$, where $\mathcal{X}$ is some base type from which (secret) values are drawn.
We write $\pi_{x}$ for the probability that the secret value is $x$.
Next, the \emph{adversary} is characterized by a set of \emph{actions} $\mathcal{W}$ and a \emph{gain function} $g : \mathcal{W} \times \mathcal{X} \to \mathbb{R}_{\geq 0}$. If the adversary performs action $w \in \mathcal{W}$ when the true value is  $x \in \mathcal{X}$, then the adversary's gain is $g(w, x)$, a non-negative real number.
Thus, the adversary's prior expected gain relative to $g$ when their uncertainty is $\pi$ is given by the maximal \emph{average vulnerability}, taken over all possible actions \cite[Def. 3.2]{Alvim2020}:
\begin{equation}
    \label{eq:prior-vulnerability}
    V_g (\pi) := \max_{w \in \mathcal{W}} \sum_{x \in \mathcal{X}} \pi_{x} g(w,x).
\end{equation}

\paragraph{Information leakage from a channel}
QIF models systems that process information as \emph{information-theoretical channels}, represented by stochastic matrices $\mathsf{C}$ that map from a (finite) set of \emph{secret inputs} $\mathcal{X}$ to a (finite) set of \emph{observable outputs} $\mathcal{Y}$, denoted $\mathsf{C} : \mathcal{X} \to \mathcal{Y}$.
We write $\mathsf{C}_{x,y}$ for the conditional probability of getting output $y$ given input $x$.
Note that QIF makes the worst-case assumption that the adversary knows the channel; this is similar to the assumption that the adversary knows the program code (as in the analysis here).

Given a channel matrix $\mathsf{C}$ representing the information leakage behavior of a system and a prior probability distribution $\pi$ on the secret, we write $[\pi \triangleright \mathsf{C}]$ for the effect of the adversary's uncertainty change after observing some output $y \in \mathcal{Y}$.
Operationally, we compute the joint distribution realized as the matrix $\mathsf{J}_{x,y} = \pi_{x} \mathsf{C}_{x,y}$ and, for each output $y$, we compute the conditional (posterior) probability distribution on the secret given $y$, representing the adversary's changed uncertainty, i.e. for each output $y$ we have a probability distribution on the secret.
Associated with each output $y$ is the marginal probability that it will be observed, and so $[\pi \triangleright \mathsf{C}]$ can be thought of as a distribution over posterior distributions, and it is sufficient to compute the posterior vulnerability.
    
\paragraph{Posterior Vulnerability}
The adversary's final expected gain allows the choice of different actions depending on their observation, weighted according to the marginal probabilities \cite[Thm. 5.7]{Alvim2020}:
\begin{equation}
    \label{eq:posterior-vulnerability}
    V_g [\pi \triangleright \mathsf{C}] := \sum_{y \in \mathcal{Y}} \max_{w \in \mathcal{W}} \sum_{x \in \mathcal{X}} \pi_{x} \mathsf{C}_{x,y} g(w,x).
\end{equation}


\paragraph{Multiplicative Leakage}
The ratio of the posterior and prior vulnerabilities measures how much an adversary has learned from observing the system output $y \in \mathcal{Y}$.
As the ratio increases, greater is the leakage from the system and more is learned by the adversary, which is defined as \cite[Def. 5.11]{Alvim2020}:
\begin{equation}
    \label{eq:leakage}
    \mathcal{L}_{g}^{\times} (\pi, C) := \frac{V_g [\pi \triangleright \mathsf{C}]}{V_g (\pi)}.
\end{equation}

We shall make use of the following channel compositions in order to simplify our information leakage models.

\paragraph{Cascading}
Given channel matrices $\mathsf{C} : \mathcal{X} \to \mathcal{Y}$ and $\mathsf{D} : \mathcal{Y} \to \mathcal{Z}$, the channel matrix $\mathsf{CD} : \mathcal{X} \to \mathcal{Z}$ is \emph{the cascading of channel $\mathsf{C}$ followed by channel $\mathsf{D}$}, where $\mathsf{CD}$ is given by matrix multiplication \cite[Def. 4.18]{Alvim2020}.
We use this concept to split both the Topics API and the third-party cookies pipelines into their respective components.

\paragraph{Internal Fixed-Probability Choice}
This composition of two compatible channel matrices, i.e. both with the same input set $\mathcal{X}$, models a scenario in which the system can (privately) choose between two possible paths according to a fixed probability $r$.
We use this to account for the differential privacy aspect of the Topics API.

Formally, \emph{the internal probabilistic choice with fixed-probability $r$ of channel matrices $\mathsf{C}^{1} : \mathcal{X} \to \mathcal{Y}^{1}$ and $\mathsf{C}^{2} : \mathcal{X} \to \mathcal{Y}^{2}$} is the channel $(\mathsf{C}^1 \ \oplus_{r}\ \mathsf{C}^2) : \mathcal{X} \to (\mathcal{Y}^{1} \cup \mathcal{Y}^{2})$ defined as \cite[Def. 8.5]{Alvim2020}:
\begin{equation}
    \label{eq:internal-choice}
    (\mathsf{C}^1 \ \oplus_{r}\ \mathsf{C}^2)_{x,y} =
    \begin{cases}
        (1-r) \mathsf{C}_{x,y}^{1} + r \mathsf{C}_{x,y}^{2} & \quad \text{if } y \in \mathcal{Y}^{1} \cap \mathcal{Y}^{2} \text{,} \\
        (1-r) \mathsf{C}_{x,y}^{1} & \quad \text{if } y \in \mathcal{Y}^{1} - \mathcal{Y}^{2} \text{,} \\
        r \mathsf{C}_{x,y}^{2} & \quad \text{if } y \in \mathcal{Y}^{2} - \mathcal{Y}^{1} \text{.}
    \end{cases}
\end{equation}

Moreover, QIF allows the computation of channel capacity, which bounds information leakage for a given channel for any prior probability distribution and gain function, using Bayes vulnerability.

\paragraph{Bayes vulnerability}
This measure returns the likelihood of an adversary correctly guessing the secret value in one try.
It is defined by the identity gain function, $g_{\mathit{id}} : \mathcal{X} \times \mathcal{X} \to \{ 0,1 \}$ \cite[Def. 3.5]{Alvim2020}:
\begin{equation}
    \label{eq:id-gain}
    g_{\mathit{id}} (w,x) =
    \begin{cases}
        1 & \quad \text{if } w = x \text{,} \\
        0 & \quad \text{if } w \neq x \text{,}
    \end{cases}
\end{equation}
where the set of possible adversary actions equals the set of possible secret values, $\mathcal{W} = \mathcal{X}$.
The prior  $V_{1} (\pi)$ (cf.~\eqref{eq:prior-vulnerability}) and the posterior $V_{1} [\pi \triangleright \mathsf{C}]$ (cf.~\eqref{eq:posterior-vulnerability}) Bayes vulnerabilities simplify respectively to:\footnote{$V_{1} (\pi) = V_{g_{\mathit{id}}} (\pi)$ and $V_{1} [\pi \triangleright \mathsf{C}] = V_{g_{\mathit{id}}} [\pi \triangleright \mathsf{C}]$, where we use the subscript $1$ instead of $g_{\mathit{id}}$ to specify the adversary's (mandatory) one attempt to guess the secret.}
\begin{equation}
    \label{eq:bayes-vulnerability-initial}
    V_{1} (\pi) = \max_{x \in \mathcal{X}} \pi_{x}~,
\end{equation}
\begin{equation}
    \label{eq:bayes-vulnerability-final}
    V_{1} [\pi \triangleright \mathsf{C}] = \sum_{y \in \mathcal{Y}} \max_{x \in \mathcal{X}} \mathsf{J}_{x,y}.
\end{equation}

\paragraph{Channel (average-case) Capacity}
The \emph{multiplicative (average-case) Bayes capacity of a channel $\mathsf{C}$} is the maximum multiplicative leakage over all gain functions and prior distributions \cite[Thm. 7.5]{Alvim2020}.
It is realized on a uniform prior $\vartheta$ for Bayes leakage \cite[Thm. 7.2]{Alvim2020}:
\begin{equation}
    \label{eq:bayes-capacity}
    \mathcal{ML}_{g}^{\times} (\mathbb{D}, \mathsf{C}) \leq \mathcal{ML}_{1}^{\times} (\mathbb{D}, \mathsf{C}) = \mathcal{L}_{1}^{\times} (\vartheta, \mathsf{C}) = \sum_{y \in \mathcal{Y}} \max_{x \in \mathcal{X}} \mathsf{C}_{x,y}.
\end{equation}

Notice that the capacity defined above scales with the size of the channel, i.e. if a channel has $M$ columns then its capacity is always at most $M$ ($\mathcal{ML}_{g}^{\times} (\mathbb{D}, \mathsf{C}) \leq M$), with the maximum capacity of $M$ occurring for the channel that leaks everything.
This can be interpreted as saying that an adversary's gain in knowledge from using the channel (as opposed to just their prior knowledge) is at most an $M$-fold gain.
For instance, for the channel that leaks everything, the adversary's worst (lowest) prior is uniform (i.e. $\nicefrac{1}{M}$) and their posterior probability of guessing the secret is $1$, hence the multiplicative gain for this adversarial scenario is exactly $M$.

The (average-case) capacity is a tight upper-bound on the \emph{average leakage} the channel can cause, quantified over all possible Bayesian adversaries' goals and capabilities (modeled as gain functions) and states of knowledge (modeled as secrets and prior distributions).
Hence, we use (average-case) channel capacity to provide bounds to privacy for both the Topics API and third-party cookies.

\paragraph{Channel (max-case) Capacity}
The concept of (max-case) channel capacity was recently introduced to QIF \cite{Fernandes2023} as the maximum multiplicative \emph{max-case} leakage over all gain functions and prior distributions (i.e. quantified over all possible Bayesian adversaries).
Hence, it provides a tight upper-bound on the \emph{maximum leakage} the channel can cause, where the maximum is taken over all possible observations, while the (average-case) channel capacity bounds the channel's \emph{expected leakage}, also over all possible observations.

For a channel $\mathsf{C}$ modeling a local differential privacy mechanism, the (max-case) channel capacity $\mathcal{ML}_{g}^{\max}(\mathsf{C})$ is exactly equal to $e^\epsilon$, where $\epsilon$ is the differential privacy parameter \cite{Fernandes2023}:
\begin{equation}
    \label{eq:lift-capacity-epsilon}
	\mathcal{ML}_{g}^{\max}(\mathsf{C}) = \max_{x, x', y ~:~ \mathsf{C}_{x',y} > 0} \frac{\mathsf{C}_{x,y}}{\mathsf{C}_{x', y}} = e^{\epsilon}.
\end{equation}

Notice that this result is only well-defined for a channel $\mathsf{C}$ which contains no $0$ probabilities, otherwise the (max-case) channel capacity is infinite, as is the value of $\epsilon$ for local differential privacy.

The (max-case) channel capacity can be interpreted as the adversary's maximal gain in knowledge over \emph{any} single observation that could be made from the channel.  
In contrast to the (average-case) channel capacity, the (max-case) channel capacity is not realized on a uniform prior distribution; instead, it occurs as the limit of a prior which is (at the limit) a point prior on a particular secret.
Hence, we use (max-case) channel capacity to account for outliers and worst-case privacy scenarios for the Topics API.

\begin{remark}
    \label{rem:capacity-prior}
    In practice, this difference in the prior distributions for the realizations of the capacities does not bear much significance, since the adversarial models for average-case and max-case also differ in the gain functions used to attain the respective capacities (i.e. they represent distinct adversary's goals).
\end{remark}

\begin{remark}
    \label{rem:}
    Channel capacities are robust measures of leakage since they quantify over all possible adversarial scenarios, and therefore do not require any adversarial assumptions aside from the use of a gain function to model the adversary's goals.
    The capacities we present here are \emph{tight}, i.e. there exists an adversary for whom the leakage of the channel matches the capacity exactly.
    However, one may obtain tighter results by (and only by) making some assumptions about the adversary's gain function or prior knowledge.
    We argue that this would weaken the overall privacy guarantees, and for this reason we favor the use of capacities for our privacy reasoning.
\end{remark}

Finally, QIF extends to analyses of unknown correlations.

\paragraph{Unknown Correlations}
Suppose a secret $\mathcal{X}$, with prior probability distribution $\pi : \mathbb{D} \mathcal{X}$, to be processed by a channel $\mathsf{C} : \mathcal{X} \to \mathcal{Y}$, is somehow correlated with another secret $\mathcal{Z}$, with prior $\rho : \mathbb{D} \mathcal{Z}$, via a joint distribution $J : \mathbb{D} (\mathcal{Z} \times \mathcal{X})$.
It happens that $\mathsf{C}$ may also leak information about $\mathcal{Z}$, called \emph{Dalenius leakage}: $\mathcal{DL}_{g}^{\times} (\mathsf{J}, \mathsf{C})$ \cite{Alvim2020}.

However, the Dalenius leakage of secret $\mathcal{Z}$ caused by channel $\mathsf{C}$ is bounded above by the channel's multiplicative (average-case) Bayes capacity.
For any channel $\mathsf{C}$, non-negative gain function $g$, and correlation matrix $\mathsf{J}$ \cite[Thm. 10.8]{Alvim2020}:\footnote{The joint distribution $J$ can be realized as a matrix $\mathsf{J}$, which can be factored into a marginal distribution $\rho : \mathbb{D} \mathcal{Z}$ and a channel $\mathsf{B} : \mathcal{Z} \to \mathcal{X}$, i.e. $\mathsf{J} : \rho \triangleright \mathsf{B}$. Hence, for any gain function $g$ on $\mathcal{Z}$, the \emph{Dalenius multiplicative $g$-leakage of $\mathsf{J}$ and $\mathsf{C}$} is \cite[Def. 10.5]{Alvim2020}: $$\mathcal{DL}_{g}^{\times} = \frac{V_{g} [\rho \triangleright \mathsf{BC}]}{V_{g} (\rho)} = \mathcal{L}_{g}^{\times} (\rho, \mathsf{BC}).$$}
\begin{equation}
    \label{eq:dalenius-capacity}
    \mathcal{DL}_{g}^{\times} (\mathsf{J}, \mathsf{C}) \leq \mathcal{ML}_{1}^{\times} (\mathbb{D}, \mathsf{C}).
\end{equation}

Notice that Dalenius capacity bounds the information leakage over all possible instantiations of side knowledge the adversary may have.
Hence, we use Eq.~\ref{eq:dalenius-capacity} to account for \emph{all} unknown correlations that may be available to an adversary and used as side knowledge, effectively avoiding \qm{closed-world} assumptions.

\section{Model}
\label{sec:models}

In this section, we construct the leakage models for the third-party cookies and the Topics API algorithms.
For privacy analyses, we consider an \emph{adversary} who wants to re-identify Internet users, either from their browsing histories (via cookies), or from their top-$s$ topics sets (via Topics API).
For utility analyses, we consider an \emph{analyst} working for an IBA company who either wants to map unique identifiers to browsing histories in order to derive users' interests (via cookies), or to map reported topics to sets of top-$s$ topics, i.e. to distinguish between real and random topics (via Topics API).

\subsection{Third-party Cookies Model}
\label{sec:models-cookies}

\begin{figure}[t]
    \centering
    \includegraphics[width=0.45\textwidth]{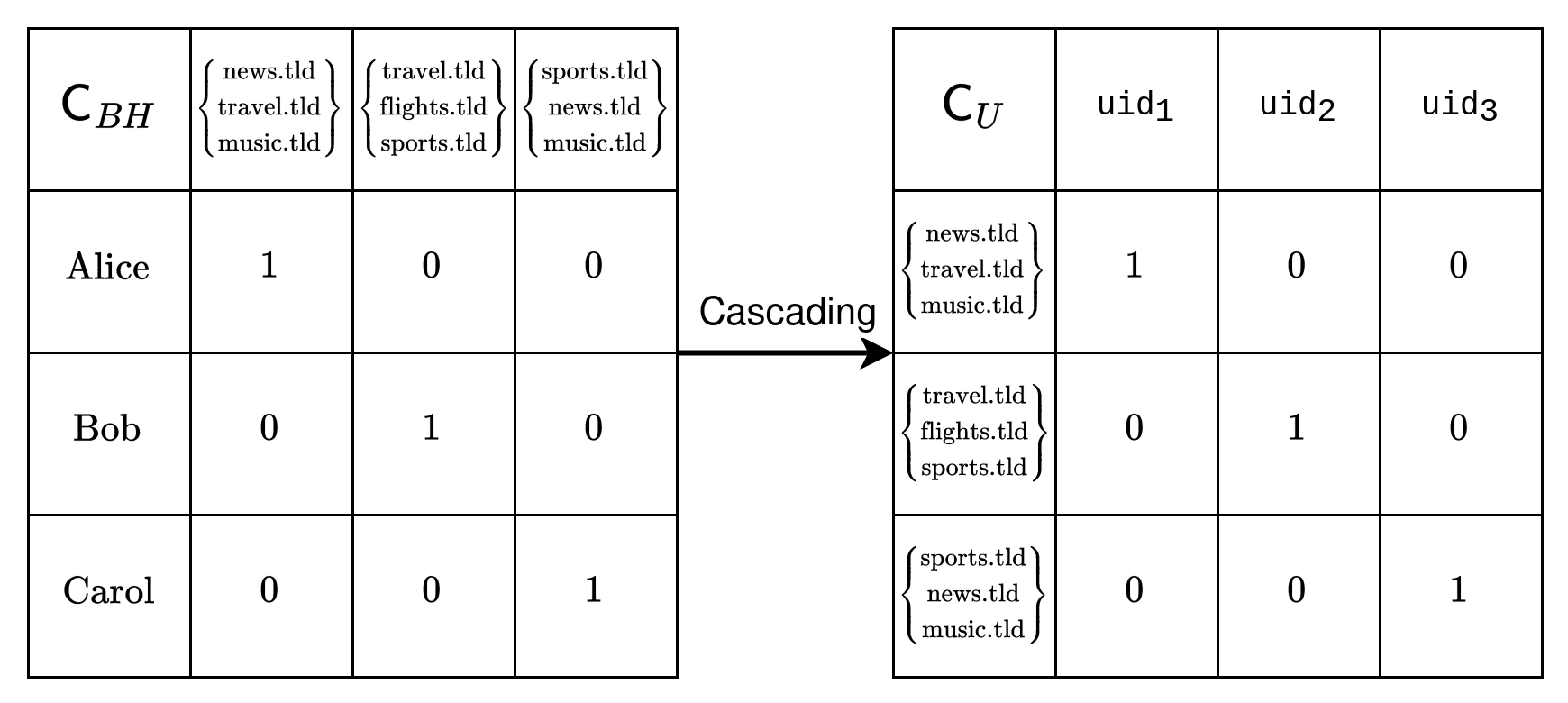}
    \caption{Pipeline for third-party cookies according to Alg.~\ref{alg:cookies}. The channel $\mathsf{C}_{\mathit{BH}}$ maps Internet users to their respective \underline{B}rowsing \underline{H}istories. The channel $\mathsf{C}_{U}$ maps browsing histories to \underline{U}nique identifiers \uid$_i$ defined by third-party cookies. The final channel for utility analyses is $\mathsf{C}_{U}$ above and the final channel for privacy analyses is $\mathsf{C}_{\mathcal{C}}$ in Fig.~\ref{fig:privacy-channels}.}
    \Description{An identity channel matrix deterministically maps Internet users to their respective browsing histories. Another identity channel matrix deterministically maps browsing histories to unique identifiers. The channels can be composed by cascading, in that order, to obtain a channel matrix that deterministically maps Internet users to unique identifiers.}
    \label{fig:cookies-channels}
\end{figure}

We assume an adversary capable of observing the whole browsing history of every Internet user, collected and reported as in Alg.~\ref{alg:cookies}.
The local storage of the browsing history occurs on line 1, while leakage of information occurs on the subsequent lines when the browser reports \cookie, \context, and \timestamp\ to a third-party.

Hence, we define the channels $\mathsf{C}_{\mathit{BH}}$, which maps Internet users to \underline{B}rowsing \underline{H}istories and corresponds to line 1 of Alg.~\ref{alg:cookies}; and $\mathsf{C}_{U}$, which maps browsing histories to \underline{U}nique identifiers \uid$_i$ defined by third-party cookies and corresponds to the subsequent lines of Alg.~\ref{alg:cookies}.
The possibility of linking \uid s to individual Internet users is modeled as the channel cascading $\mathsf{C}_{\mathit{BH}} \mathsf{C}_{U}$, e.g. Fig.~\ref{fig:cookies-channels}.

The adversary has full visibility of channel $\mathsf{C}_{U}$ and may be able to reconstruct channel $\mathsf{C}_{\mathit{BH}}$ from auxiliary information, e.g. if users are logged into online services or from gathering publicly available information \cite{Barbaro2006}.
We further assume the adversary is capable of reconstructing channel $\mathsf{C}_{\mathit{BH}}$ and performing the channel cascading, which results in channel $\mathsf{C}_{\mathcal{C}}$, e.g. Fig.~\ref{fig:privacy-channels}.

Such a powerful adversary is in line with the literature.
For instance, nearly $90\%$ of the $500$ most popular websites in 2011 included at least one third-party known for tracking users' browsing histories, with the most common at the time present on almost $40\%$ \cite{Roesner2012}.
Moreover, when considering the possibility of collusion among third-parties, the top $10$ advertising and analytics companies in 2018 could observe more than $90\%$ of users' browsing histories \cite{Bashir2018}.

Also note that both $\mathsf{C}_{\mathit{BH}}$ and $\mathsf{C}_{U}$ are \emph{identity channels}, i.e. they leak everything, since third-party cookies are persistent cross-site unique identifiers.
From the literature, we know that browsing histories are mostly unique, e.g. Olejnik et al. have shown in 2014 that $97\%$ of browsing histories with at least four visited websites are unique \cite{Olejnik2014}, a study replicated by Bird et al. in 2020 \cite{Bird2020}.

Finally, given the capabilities and intentions of the adversary and the analyst, we use channel $\mathsf{C}_{\mathcal{C}}$ for privacy analyses, e.g. Fig.~\ref{fig:privacy-channels}, and channel $\mathsf{C}_{U}$ for utility analyses, e.g. Fig.~\ref{fig:cookies-channels}, respectively.

\subsection{Topics API Model}
\label{sec:models-topics}

\begin{figure*}[t]
    \centering
    \includegraphics[width=0.95\textwidth]{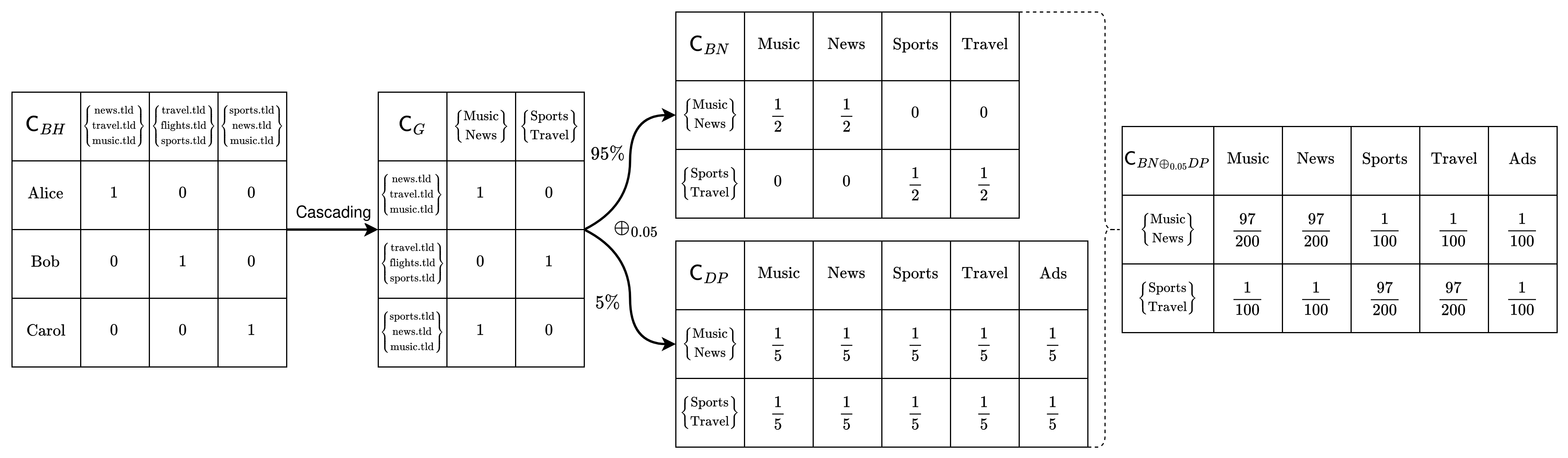}
    \caption{Pipeline for the Topics API according to Alg.~\ref{alg:topics-derive} and Alg.~\ref{alg:topics-report}. The channel $\mathsf{C}_{\mathit{BH}}$ maps Internet users to their respective \underline{B}rowsing \underline{H}istories. The channel $\mathsf{C}_{G}$ maps browsing histories to sets of top-$(s=2)$ topics, a \underline{G}eneralization step. The channel $\mathsf{C}_{\mathit{BN} \oplus_{0.05} \mathit{DP}}$ maps sets of top-$(s=2)$ topics to individual topics; this channel is the result of an internal fixed-probability choice between channels $\mathsf{C}_{\mathit{BN}}$ and $\mathsf{C}_{\mathit{DP}}$. The channel $\mathsf{C}_{\mathit{BN}}$ is the case in which the API reports a topic from the sets of top-$(s=2)$ topics with uniform probability, i.e. each with $\nicefrac{1}{s}$ probability, a \underline{B}ounded \underline{N}oise step that happens with $(1-r)=95\%$ chance. The channel $\mathsf{C}_{\mathit{DP}}$ is the case in which the API reports a random topic from the whole taxonomy with uniform probability, i.e. each with $\nicefrac{1}{m}$ probability, a \underline{D}ifferential \underline{P}rivacy step that happens with $r=5\%$ chance. The final channel for utility analyses is $\mathsf{C}_{\mathit{BN} \oplus_{0.05} \mathit{DP}}$ above and the final channel for privacy analyses is $\mathsf{C}_{\mathcal{T}}$ in Fig.~\ref{fig:privacy-channels}.}
    \Description{An identity channel matrix deterministically maps Internet users to their respective browsing histories. Another deterministic, but not necessarily identity channel matrix maps browsing histories to sets of topics of size 2. With 95 percent chance, a probabilistic channel that uniformly maps sets of topics to individual topics in the corresponding sets occurs, i.e. either 0 or 50 percent chance of a topic being in a given set of size 2. With 5 percent chance, another probabilistic channel that uniformly maps sets of topics to every possible individual topic occurs, i.e. 20 percent chance for any of the 5 topics being in a given set. The first two deterministic channels can be composed by cascading, in that order, to obtain a channel matrix that deterministically maps Internet users to sets of topics. The last two probabilistic channels can be composed by internal fixed-probability choice to obtain a probabilistic channel that maps sets of topics to every possible individual topic. The first two deterministic channels can also be composed by cascading with the probabilistic one resulting from the internal fixed-probability choice, in that order, to obtain a probabilistic channel matrix that maps Internet users to every possible individual topic.}
    \label{fig:topics-channels}
\end{figure*}

\begin{figure}[t]
    \centering
    \includegraphics[width=0.45\textwidth]{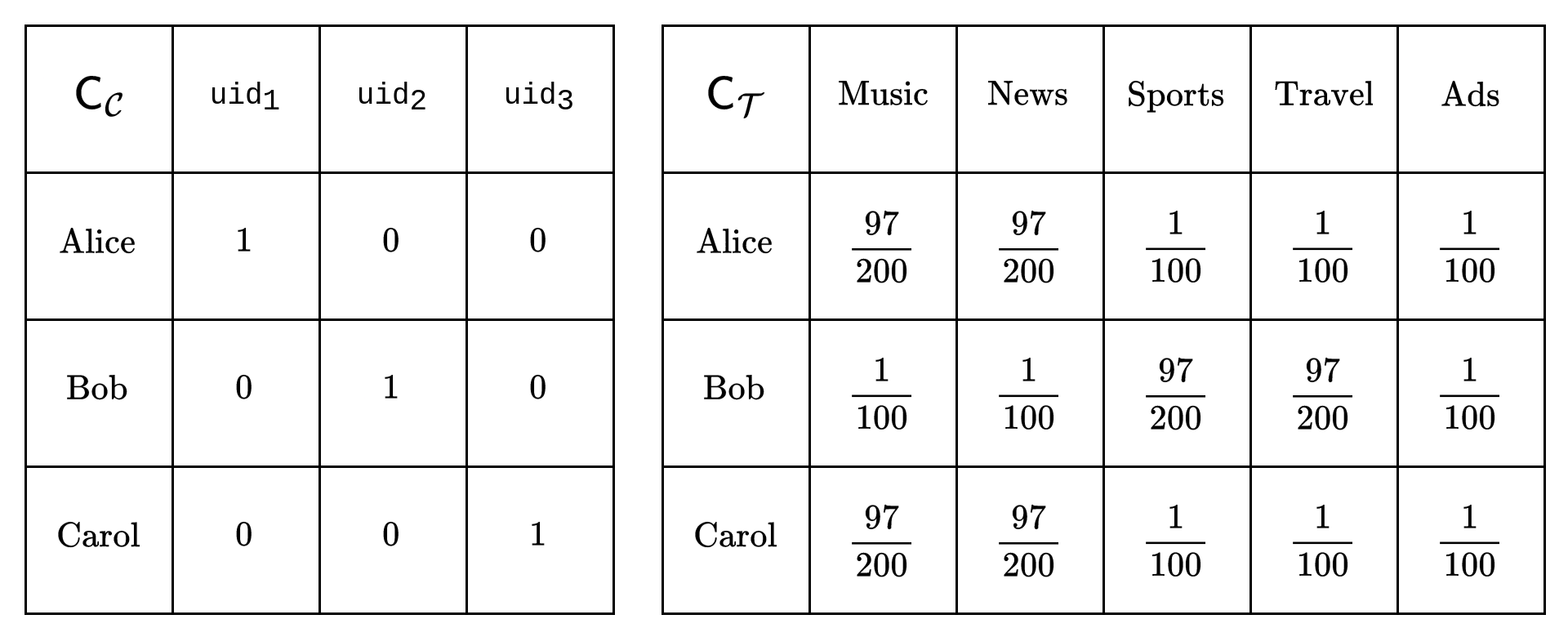}
    \caption{Final channels for privacy analyses. The channel $\mathsf{C}_{\mathcal{C}}$ is the cascading of channels $\mathsf{C}_{\mathit{BH}} \mathsf{C}_{U}$ and the channel $\mathsf{C}_{\mathcal{T}}$ is the cascading of channels $\mathsf{C}_{\mathit{BH}} \mathsf{C}_{G} \mathsf{C}_{\mathit{BN} \oplus_{0.05} \mathit{DP}}$.}
    \Description{An identity channel matrix deterministically maps Internet users to unique identifiers. Another probabilistic channel matrix maps Internet users to every possible individual topic.}
    \label{fig:privacy-channels}
\end{figure}

We assume an adversary capable of observing the whole set of top-$s$ topics of every Internet user, collected as in Alg.~\ref{alg:topics-derive} and reported as in Alg. \ref{alg:topics-report}.
We assume the browsing history used in Alg.~\ref{alg:topics-derive} was already collected during each epoch as on line 1 of Alg.~\ref{alg:cookies}.

The classification of the browsing history entries, i.e. the mapping from \context\ to $\mathit{topic}$, occurs on lines 1 and 2 of Alg.~\ref{alg:topics-derive}, while the top-$s$ topics set is computed on line 3.
The random, uniform choice of a topic from the top-$s$ set occurs on line 1 of Alg.~\ref{alg:topics-report}, while lines 2 and 3 account for the possibility of that topic being overridden by a topic chosen uniformly at random from the whole taxonomy.
The information leakage for the whole Topics API pipeline occurs on line 4 of Alg.~\ref{alg:topics-report} when the browser reports a $\mathit{topic}$ to a third-party.
We will also consider intermediate scenarios to analyze the leakage throughout the Topics API pipeline, i.e. just after line 1 of Alg.~\ref{alg:topics-report} with the reporting of a $\mathit{topic}$, and on line 4 of Alg.~\ref{alg:topics-derive} with the reporting of the whole top-$s$ set.

Hence, we define the channels $\mathsf{C}_{\mathit{BH}}$, an identity channel which maps Internet users to browsing histories and corresponds to line 1 of Alg.~\ref{alg:cookies}; $\mathsf{C}_{G}$, a deterministic channel which maps browsing histories to top-$s$ sets and corresponds to lines 1--3 of Alg.~\ref{alg:topics-derive};\footnote{We assume every user has enough browsing history entries to build their top-$s$ set.} and both $\mathsf{C}_{\mathit{BN}}$ and $\mathsf{C}_{\mathit{BN} \oplus_{r} \mathit{DP}}$, which map top-$s$ sets to individual topics and correspond to line 1 and to lines 2--3 of Alg.~\ref{alg:topics-report}, respectively.
The possibility of linking $\mathit{topic}$s or top-$s$ sets to individual users is modeled as the channel cascading $\mathsf{C}_{\mathit{BH}} \mathsf{C}_{G} \mathsf{C}_{\mathit{BN} \oplus_{r} \mathit{DP}}$, e.g. Fig.~\ref{fig:topics-channels}.

The adversary has full visibility of channels $\mathsf{C}_{G}$ and $\mathsf{C}_{\mathit{BN} \oplus_{r} \mathit{DP}}$ and may be able to reconstruct channel $\mathsf{C}_{\mathit{BH}}$ from auxiliary information, as before.
We further assume the adversary is capable of reconstructing channel $\mathsf{C}_{\mathit{BH}}$ and performing the channel cascading, which results in channel $\mathsf{C}_{\mathcal{T}}$, e.g. Fig.~\ref{fig:privacy-channels}.
Such a powerful adversary is based on the possibility of collusion among Topics API callers to link the identity of Internet users \cite{Carey2023}.

The channel $\mathsf{C}_{G}$ represents a \underline{G}eneralization step that reduces the amount of information leaked by mapping domains to sets of top-$s$ topics from a pre-defined taxonomy, introducing $k$-anonymity to the Topics API pipeline.
The value of the integer $k \geq 1$ can be derived from the channel $\mathsf{C}_{G}$ by taking the minimum count of the number $1$ on each column,\footnote{Note that the value of $k$ is not directly determined by $s$ because it is data-dependent.} e.g. $k=1$ in Fig.~\ref{fig:topics-channels}.\footnote{If $k=1$, then there is at least one top-$s$ set that can be uniquely remapped to a corresponding unique browsing history, e.g. \{Sports, Travel\} is uniquely remapped to \{travel.tld, flights.tld, sports.tld\} and uniquely identifies the user Bob in Fig.~\ref{fig:topics-channels}.}

The channel $\mathsf{C}_{\mathit{BN} \oplus_{r} \mathit{DP}}$ introduces two privacy methods to the Topics API pipeline.
With $(1-r)\%$ chance, the channel $\mathsf{C}_{\mathit{BN}}$ is executed and introduces \underline{B}ounded \underline{N}oise with parameter $B=s$, the size of the top-$s$ sets.
The value of the integer $B \geq 1$ can be derived from the channel $\mathsf{C}_{\mathit{BN}}$ by either the number of non-zero values on each row, or the denominator value on the ratios, e.g. $B=2$ in Fig.~\ref{fig:topics-channels}.
With $r\%$ chance, channel $\mathsf{C}_{\mathit{DP}}$ is executed and introduces \underline{D}ifferential \underline{P}rivacy with uniform probability of reporting any topic from the whole taxonomy, i.e. $\nicefrac{1}{m}$ for every conditional probability, where $m$ is the total number of topics in the taxonomy, e.g. $m=5$ and $r=0.05$ in Fig.~\ref{fig:topics-channels}.
This introduces the \emph{plausible deniability} property from differential privacy to the Topics API.

The channel $\mathsf{C}_{\mathit{BN} \oplus_{r} \mathit{DP}}$ is then derived from the internal fixed-probability choice between channels $\mathsf{C}_{\mathit{BN}}$ and $\mathsf{C}_{\mathit{DP}}$ (cf.~\eqref{eq:internal-choice}):
\begin{equation}
    \label{eq:model-topics-utility-channel}
    \mathsf{C}_{\mathit{BN} \oplus_{r} \mathit{DP}_{(\sigma,t)}} = 
    \begin{cases}
        \frac{1-r}{s} + \frac{r}{m} & \quad \text{if } t \in \sigma \text{,} \\
        \frac{r}{m} & \quad \text{if } t \notin \sigma \text{,}
    \end{cases}
\end{equation}
where $\sigma \in \Sigma$ is a top-$s$ set given as an input to the channel, $t \in \mathtt{T}$ is a topic from the taxonomy, $m = |\mathtt{T}|$ is the total number of topics in the taxonomy, $s = |\sigma|$ is the size of the top-$s$ sets, and $r$ is the probability of the Topics API returning a random topic from the whole taxonomy instead of a random topic from a user's top-$s$ set.

Finally, given the capabilities and intentions of the adversary and the analyst, we use channel $\mathsf{C}_{\mathcal{T}}$ for privacy analyses, e.g. Fig.~\ref{fig:privacy-channels}, and channel $\mathsf{C}_{\mathit{BN} \oplus_{r} \mathit{DP}}$ for utility analyses, e.g. Fig.~\ref{fig:cookies-channels}, respectively.

\section{Theoretical Evaluation}
\label{sec:theoretical}

We evaluate the privacy vulnerability for individuals in Sec.~\ref{sec:theoretical-privacy} and the utility for IBA companies in Sec.~\ref{sec:theoretical-utility}.

\subsection{Privacy Vulnerability for Individuals}
\label{sec:theoretical-privacy}

We assume a uniform prior probability distribution over all Internet users, i.e. an adversary that has no prior knowledge of correlations between the secrets (individuals' identities) and the observables (\uid s or reported topics).
We assess privacy using the Bayes vulnerability measure, i.e. how likely an adversary is to correctly guess the value of the secret in one try.
We recall that the multiplicative (average-case) Bayes capacity is always realized on a uniform prior (cf.~\eqref{eq:bayes-capacity}) and that it is the maximum multiplicative leakage over all gain functions and prior distributions (cf.~\eqref{eq:bayes-capacity}), including the leakage for unknown, arbitrary correlations (cf.~\eqref{eq:dalenius-capacity}).
Hence, the prior privacy vulnerability is (cf.~\eqref{eq:bayes-vulnerability-initial}):
\begin{equation}
    \label{eq:theoretical-privacy-vulnerability-initial}
    V_1 (\pi) = V_1 (\vartheta) = \frac{1}{N},
\end{equation}
where $N = |\mathcal{X}|$ is the total number of Internet users.
For instance, when $N = 3$ in Fig.~\ref{fig:privacy-channels}, we have $V_1 (\vartheta) = \nicefrac{1}{3}$ for both third-party cookies and the Topics API.

\subsubsection{Third-party Cookies (average-case) Leakage}
\label{sec:theoretical-privacy-cookies}

Considering the model described in Sec.~\ref{sec:models-cookies}, the deterministic channel $\mathsf{C}_{\mathcal{C}}$, e.g. Fig.~\ref{fig:privacy-channels}, the adversary's uniform prior probability distribution, and the Bayes vulnerability measure, the posterior privacy vulnerability is:
\begin{equation}
    \label{eq:theoretical-privacy-cookies-vulnerability-final}
    V_1 [\vartheta \triangleright \mathsf{C}_{\mathcal{C}}] = \frac{1}{N} \sum_{h=2}^{h=c} \binom{c}{h} = \frac{1}{N} \sum_{h=2}^{h=c} \frac{c!}{h! (c - h)!},
\end{equation}
where $N = |\mathcal{X}|$ is the total number of Internet users, $c$ is the number of contexts on the Internet that include third-party cookies, and $h$ is the number of contexts in a user's browsing history that may be affected by third-party cookies cross-site tracking.
Hence, the (average-case) leakage caused by third-party cookies is:
\begin{equation}
    \label{eq:theoretical-privacy-cookies-leakage}
    \mathcal{L}^{\times}_1 (\vartheta, \mathsf{C}_{\mathcal{C}}) = \sum_{h=2}^{h=c} \frac{c!}{h! (c - h)!} = \min\{N, 2^{c} - c - 1\},
\end{equation}
i.e. the channel $\mathsf{C}_{\mathcal{C}}$ leaks everything.\footnote{Note that by considering the number of contexts on the Internet and in a user's browsing history, we are actually using channel $\mathsf{C}_{\mathit{BH}}$ to compute the results above. This is possible since $\mathsf{C}_{\mathit{BH}}$, $\mathsf{C}_{U}$, and $\mathsf{C}_{\mathcal{C}}$ are all identity channels, hence browsing histories and unique identifiers are essentially the same from the adversary's perspective.}
For instance, when $N = 3$ and $c = 5$ in Fig.~\ref{fig:privacy-channels}, we have $V_1 [\vartheta \triangleright \mathsf{C}_{\mathcal{C}}] = \nicefrac{26}{3}$ and $\mathcal{L}^{\times}_1 (\vartheta, \mathsf{C}_{\mathcal{C}}) = 26$.

\subsubsection{Topics API (average-case) Leakage}
\label{sec:theoretical-privacy-topics-average}

Considering the model described in Sec.~\ref{sec:models-topics} and the Bayes vulnerability measure, we analyze the Topics API leakage incrementally for each stage of the execution sequence of its algorithms, i.e. we consider the corresponding channel cascading to describe the respective leakage.

\paragraph{Generalization (Alg.~\ref{alg:topics-derive})}
This method was first formalized for privacy applications by Samarati and Sweeney as a way of achieving $k$-anonymity \cite{Samarati1998}.
When applied, it replaces an attribute's values by a more generic one, e.g. topics instead of domains.
The addition of the channel $\mathsf{C}_{G}$ (lines 1--3 of Alg.~\ref{alg:topics-derive}) to the pipeline introduces generalization and reduces the amount of information reported by the Topics API by decreasing the number of possible outputs, when compared to reporting users' complete browsing histories.

Considering the deterministic channel cascading $\mathsf{C}_{\mathit{BH}} \mathsf{C}_{G}$, e.g. Fig.~\ref{fig:topics-channels}, and the adversary's uniform prior probability distribution, the posterior privacy vulnerability up to the generalization step is:
\begin{equation}
    \label{eq:theoretical-privacy-topics-average-generalization-vulnerability-final}
    V_1 [\vartheta \triangleright (\mathsf{C}_{\mathit{BH}} \mathsf{C}_{G})] = \sum_{\sigma \in \Sigma} \max_{x \in \mathcal{X}} \mathsf{J}_{x, \sigma} = \frac{|\Sigma|}{N} \leq \frac{1}{N} \binom{m}{s},
\end{equation}
where $\Sigma$ is the set of top-$s$ sets that occur, i.e. $|\Sigma|$ is the number of columns on the channel $\mathsf{C}_{\mathit{BH}} \mathsf{C}_{G}$, $\mathcal{X}$ is the set of Internet users, $\mathsf{J}$ is the joint matrix,\footnote{Given a prior probability distribution $\pi : \mathbb{D} \mathcal{X}$ and a channel $\mathsf{C} : \mathcal{X} \to \mathcal{Y}$, the \emph{joint matrix} $\mathsf{J}$ is the multiplication of each row of $\mathsf{C}$ by the corresponding probability $\pi_x$.} $m$ is the number of topics in the taxonomy, and $s = |\sigma|$ is the number of topics in the top-$s$ sets.
Hence, the Topics API (average-case) leakage up to the generalization step is:
\begin{equation}
    \label{eq:theoretical-privacy-topics-average-generalization-leakage}
    \mathcal{L}^{\times}_1 (\vartheta, \mathsf{C}_{\mathit{BH}} \mathsf{C}_{G}) = |\Sigma| \leq \binom{m}{s}.
\end{equation}
For instance, when $N = 3$, $|\Sigma| = 2$, and $m = 4$ in Fig.~\ref{fig:topics-channels}, we have $V_1 [\vartheta \triangleright (\mathsf{C}_{\mathit{BH}} \mathsf{C}_{G})] = \nicefrac{2}{3} \leq 3$ and $\mathcal{L}^{\times}_1 (\vartheta, \mathsf{C}_{\mathit{BH}} \mathsf{C}_{G}) = 2 \leq 6$.

\paragraph{Bounded Noise (Alg.~\ref{alg:topics-report})}
This method introduces uncertainty by randomizing the true value of a secret across a bounded set.
When applied, it allows some of the previously impossible output values to occur.
In QIF, this is modeled as a stochastic, non-deterministic channel that contains one or more zeroes.
The addition of the channel $\mathsf{C}_{\mathit{BN}}$ (line 1 of Alg.~\ref{alg:topics-report}) to the pipeline introduces bounded noise and changes the output from top-$s$ sets to individual topics, further reducing the amount of information reported by the API.

Considering the channel cascading $\mathsf{C}_{\mathit{BH}} \mathsf{C}_{G} \mathsf{C}_{\mathit{BN}}$, e.g. Fig.~\ref{fig:topics-channels}, and the adversary's uniform prior probability distribution, the posterior privacy vulnerability up to the bounded noise step is:
\begin{equation}
    \label{eq:theoretical-privacy-topics-average-bounded-noise-vulnerability-final}
    V_1 [\vartheta \triangleright (\mathsf{C}_{\mathit{BH}} \mathsf{C}_{G} \mathsf{C}_{\mathit{BN}})] = \frac{1}{N} \frac{m'}{s} \leq \frac{1}{N} \frac{m}{s},
\end{equation}
where $m'$ is the number of topics that occur, i.e. the number of columns on the channel $\mathsf{C}_{\mathit{BH}} \mathsf{C}_{G} \mathsf{C}_{\mathit{BN}}$, $m$ is the number of topics in the taxonomy, and $s$ is the number of topics in the top-$s$ sets.
Hence, the Topics API (average-case) leakage up to the bounded noise step is:
\begin{equation}
    \label{eq:theoretical-privacy-topics-average-bounded-noise-leakage}
    \mathcal{L}^{\times}_1 (\vartheta, \mathsf{C}_{\mathit{BH}} \mathsf{C}_{G} \mathsf{C}_{\mathit{BN}}) = \frac{m'}{s} \leq \frac{m}{s}.
\end{equation}
For instance, when $N = 3$, $m' = 3$, $m = 4$, and $s = 2$ in Fig.~\ref{fig:topics-channels}, $V_1 [\vartheta \triangleright (\mathsf{C}_{\mathit{BH}} \mathsf{C}_{G} \mathsf{C}_{\mathit{BN}})] = \nicefrac{1}{2} \leq \nicefrac{2}{3}$, $\mathcal{L}^{\times}_1 (\vartheta, \mathsf{C}_{\mathit{BH}} \mathsf{C}_{G} \mathsf{C}_{\mathit{BN}}) = \nicefrac{3}{2} \leq 2$.

\paragraph{Differential Privacy (Alg.~\ref{alg:topics-report})}
This method was first formalized for privacy applications by Dwork et al. as a measure of indistinguishability of secrets in adjacent datasets \cite{Dwork2006}.
It was later extended to account for the adjacency of individual records \cite{Kasiviswanathan2008} and for general indistinguishability metrics \cite{Chatzikokolakis2013,Fernandes2021}.
A channel $\mathsf{C} : \mathcal{X} \to \mathcal{Y}$ satisfies $\epsilon$-differential-privacy ($\epsilon$-DP) if distributions $\Delta_{1,2}$ on $\mathcal{Y}$ that result from the channel's execution on adjacent inputs $x_{1,2} \in \mathcal{X}$, i.e. $\mathsf{C} (x_1)$ and $\mathsf{C} (x_2)$, are bounded by the parameter $\epsilon$ for all subsets $Y \in \mathcal{Y}$:\footnote{The definition of adjacency depends on $\mathcal{X}$ and on the notion of indistinguishability. In the original formulation of $\epsilon$-DP, \emph{central differential privacy} \cite{Dwork2006}, $\mathcal{X}$ contains microdata datasets, e.g. individuals' data on each row, and two datasets $x_{1,2}$ are adjacent if they differ in the presence or absence of a single row, i.e. it protects against membership inference attacks. In \emph{local differential privacy} \cite{Kasiviswanathan2008}, $\mathcal{X}$ contains single values (or rows) and every pair $x_{1,2}$ is considered adjacent, i.e. it protects against attribute inference attacks. Because $\epsilon$-DP for a channel $\mathsf{C}$ depends only on the input data $\mathcal{X}$ and its adjacency relationship, but not on the prior distribution on $\mathcal{X}$, differential privacy is said to be independent of prior knowledge.}
\begin{equation}
    \label{eq:epsilon-differential-privacy}
    \left| \ln \frac{\Delta_{1} (Y)}{\Delta_{2} (Y)} \right| \leq \epsilon.
\end{equation}
The smaller the value of $\epsilon$, the more private the channel is said to be.
If the distributions $\Delta_{1,2}$ are discrete, then Eq.~\ref{eq:epsilon-differential-privacy} can be simplified to the ratio of the probabilities $p_{1,2}$ for all $y \in \mathcal{Y}$ by $\mathsf{C} (x_{1,2})$.
We will consider only discrete distributions here.

The addition of the channel $\mathsf{C}_{\mathit{DP}}$ to the pipeline as an internal choice option together with channel $\mathsf{C}_{\mathit{BN}}$, i.e. as channel $\mathsf{C}_{\mathit{BN} \oplus_{r} \mathit{DP}}$ (lines 2 and 3 of Alg.~\ref{alg:topics-report}), introduces differential privacy and its plausible deniability property and completes the Topics API pipeline.

Considering the channel cascading $\mathsf{C}_{\mathit{BH}} \mathsf{C}_{G} \mathsf{C}_{\mathit{BN} \oplus_{r} \mathit{DP}}$, e.g. Fig.~\ref{fig:topics-channels}, which results in channel $\mathsf{C}_{\mathcal{T}}$, e.g. Fig.~\ref{fig:privacy-channels}, and the adversary's uniform prior probability distribution, the posterior privacy vulnerability of the complete Topics API is:
\begin{align}
    \label{eq:theoretical-privacy-topics-average-complete-vulnerability-final}
    &V_1 [\vartheta \triangleright (\mathsf{C}_{\mathit{BH}} \mathsf{C}_{G} \mathsf{C}_{\mathit{BN} \oplus_{r} \mathit{DP}})] =\nonumber\\
    &= V_1 [\vartheta \triangleright \mathsf{C}_{\mathcal{T}}] = \frac{1}{N} \left( r + \frac{m'(1-r)}{s} \right) \leq \frac{1}{N} \left( r + \frac{m(1-r)}{s} \right),
\end{align}
where $m'$ is the number of topics that occur, i.e. the number of columns on the channel $\mathsf{C}_{\mathcal{T}}$, $m$ is the number of topics in the taxonomy, $r$ is the probability that the Topics API reports a random topic from the whole taxonomy, and $s$ is the number of topics in the top-$s$ sets.
Hence, the complete Topics API (average-case) leakage is:
\begin{align}
    \label{eq:theoretical-privacy-topics-average-complete-leakage}
    &\mathcal{L}_{1}^{\times} (\vartheta, \mathsf{C}_{\mathit{BH}} \mathsf{C}_{G} \mathsf{C}_{\mathit{BN} \oplus_{r} \mathit{DP}}) =\nonumber\\
    &\mathcal{L}_{1}^{\times} (\vartheta, \mathsf{C}_{\mathcal{T}}) = r + \frac{m' (1 - r)}{s} \leq r + \frac{m (1 - r)}{s}.
\end{align}
For instance, when $N = 3$, $m' = 3$, $m = 4$, $s = 2$, and $r = 0.05$ in Fig.~\ref{fig:topics-channels}, $V_1 [\vartheta \triangleright \mathsf{C}_{\mathcal{T}}] = 0.65 \leq 0.81$ and $\mathcal{L}_{1}^{\times} (\vartheta, \mathsf{C}_{\mathcal{T}}) = 1.95 \leq 2.43$.

Additionally, considering the discrete metric for indistinguishability \cite{Fernandes2021},\footnote{Roughly, in metric differential privacy \cite{Chatzikokolakis2013}, a system $C : \mathcal{X} \to \mathbb{D} \mathcal{Y}$ satisfies ${\epsilon}{\cdot}{\mathbf{d}}$-privacy iff $\forall x,x' \in \mathcal{X}$, $C(x) \leq e^{\epsilon \cdot \mathbf{d}(x,x')} C(x')$. For the discrete metric, $d(x,x') = 1$ whenever $x \neq x'$} the value of the parameter $\epsilon$ for the Topics API can be obtained from the channel $\mathsf{C}_{\mathit{BN} \oplus_{r} \mathit{DP}}$:
\begin{equation}
    \label{eq:theoretical-privacy-topics-epsilon}
    \epsilon_{\mathsf{C}_{\mathcal{T}}} = \epsilon_{\mathsf{C}_{\mathit{BN} \oplus_{r} \mathit{DP}}} = \ln \left( 1 + \frac{m(1{-}r)}{rs} \right).
\end{equation}

\subsubsection{Topics API (max-case) Capacity}
\label{sec:theoretical-privacy-topics-max}

We recall that all the leakage results from Sec.~\ref{sec:theoretical-privacy-cookies} and Sec.~\ref{sec:theoretical-privacy-topics-average} are also the channel (average-case) capacities for their respective channels (cf.~\eqref{eq:bayes-capacity}).

Given the value of the parameter $\epsilon$, we can also compute the channel (max-case) capacity for a local differential privacy channel, e.g. $\mathsf{C}_{\mathit{BN} \oplus_{r} \mathit{DP}}$, (cf.~\eqref{eq:lift-capacity-epsilon}): $\mathcal{ML}_{g}^{\max} (\mathsf{C}) = e^{\epsilon}$.
Hence, the Topics API (max-case) capacity is:
\begin{equation}
    \label{eq:theoretical-privacy-topics-max-complete-capacity}
    \mathcal{ML}_{g}^{\max} (\mathsf{C}_{\mathcal{T}}) = \mathcal{ML}_{g}^{\max} (\mathsf{C}_{\mathit{BN} \oplus_{r} \mathit{DP}}) = 1 + \frac{m(1-r)}{rs}.
\end{equation}
For instance, when $N = 3$, $m = 4$, $s = 2$, and $r = 0.05$ in Fig.~\ref{fig:topics-channels}, we have $\epsilon_{\mathsf{C}_{\mathcal{T}}} = 3.6636$ and $\mathcal{ML}_{g}^{\max} (\mathsf{C}_{\mathcal{T}}) = 39$.

\subsection{Utility for IBA Companies}
\label{sec:theoretical-utility}

In this section, we consider the accuracy of some inferences that the (interest-based advertising) analyst would like to make.
We can also use QIF in this scenario by defining precise gain functions that describe the analyst's goals.
In Sec.~\ref{sec:theoretical-utility-counting}, the analyst wants to learn the most popular topic from all the reported topics.
In Sec.~\ref{sec:theoretical-utility-iba-gain}, the analyst wants to determine the authenticity of a received topic, i.e. whether it is a genuine topic or a randomly generated one.

\subsubsection{Learning the Most Popular Topic}
\label{sec:theoretical-utility-counting}

The analyst wants to learn the most popular topic from all the reported topics by the users.
Hence, we develop a model for the aggregate set of noisy reported results represented by an aggregate channel from all individual channels using the Kronecker product, i.e. we consider each user has a corresponding channel $\mathsf{C}_{\mathit{BN} \oplus_{r} \mathit{DP}}$ that leaks information on whether the reported topic is genuine or not, and their Kronecker product is the channel corresponding to the aggregate information received by the analyst.
The analyst then post-processes the aggregate channel with a query that reports the total (noisy) counts, and the utility is measured as a function of the real count, i.e. as a gain function that returns $1$ if the analyst's count equals the real count.

\paragraph{Kronecker Product}
It has been shown that Kronecker products can be used to model the aggregate utility of complex datasets derived from multiple individuals in a local differential privacy context \cite[Sec. 4.2]{Alvim2023}.
Moreover, for random response mechanisms, e.g. the channel $\mathsf{C}_{\mathit{BN} \oplus_{r} \mathit{DP}}$, post-processed by a counting query, the utility is stable with respect to $\epsilon$, i.e. monotonic \cite[Sec. 6.1]{Alvim2023}.

\begin{definition}
    \label{def:kronecker-product}
    Given real-valued matrices $\mathsf{A} : \mathcal{X} \to \mathcal{Y}$ and $\mathsf{A'} : \mathcal{X'} \to \mathcal{Y'}$, the \emph{Kronecker product $\mathsf{A} \otimes \mathsf{A'} : (\mathcal{X}, \mathcal{X'}) \to (\mathcal{Y}, \mathcal{Y'})$} is defined as:
    \begin{equation}
        \label{eq:kronecker-product}
        (\mathsf{A} \otimes \mathsf{A'})_{(x,x'),(y,y')} := \mathsf{A}_{x,y} \cdot \mathsf{A'}_{x',y'}.
    \end{equation}
    Moreover, we write $\mathsf{A}^{\otimes N}$ for the $N$-fold product of $\mathsf{A}$ with itself, i.e. $\mathsf{A}^{\otimes N} = \mathsf{A}^{\otimes N-1} \otimes \mathsf{A}$ and $\mathsf{A}^{0}$ is the $1 \times 1$ identity $\mathbb{I}$.
\end{definition}

Using the Kronecker product, we are able to derive the series of novel results that we present in the remaining of this subsection.

\begin{theorem}
    \label{theo:kronecker-capacity}
    For channel matrices $\mathsf{C} : \mathcal{X} \to \mathcal{Y}$ and $\mathsf{D} : \mathcal{X'} \to \mathcal{Y'}$, the multiplicative (average-case) Bayes capacity of the Kronecker product $\mathsf{C} \otimes \mathsf{D} : (\mathcal{X}, \mathcal{X'}) \to (\mathcal{Y}, \mathcal{Y'})$ is:
    \begin{equation*}
        \mathcal{ML}_{1}^{\times} (\vartheta'', \mathsf{C} \otimes \mathsf{D}) = \mathcal{ML}_{1}^{\times} (\vartheta, \mathsf{C}) \cdot \mathcal{ML}_{1}^{\times} (\vartheta', \mathsf{D}),
    \end{equation*}
    i.e. the product of the multiplicative Bayes capacities of $\mathsf{C}$ and $\mathsf{D}$.
\end{theorem}

\begin{corollary}
    \label{cor:kronecker-vulnerability}
    The Bayes vulnerability of the Kronecker product of channel matrices $\mathsf{C} : \mathcal{X} \to \mathcal{Y}$ and $\mathsf{D} : \mathcal{X'} \to \mathcal{Y'}$ is:
    \begin{equation*}
        V_1 [\vartheta'' \triangleright (\mathsf{C} \otimes \mathsf{D})] = V_1 [\vartheta \triangleright \mathsf{C}] \cdot V_1 [\vartheta' \triangleright \mathsf{D}],
    \end{equation*}
    i.e. the product of the Bayes vulnerabilities of $\mathsf{C}$ and $\mathsf{D}$.
\end{corollary}

\paragraph{Counting Experiment}
Considering the channel $\mathsf{C}_{\mathit{BN} \oplus_{r} \mathit{DP}}$ from Eq.~\ref{eq:model-topics-utility-channel}, we define $\mathtt{p} = \nicefrac{1-r}{s} + \nicefrac{r}{m}$, $\mathtt{q} = \nicefrac{r}{m}$, and the channel $\mathsf{C}_{\in}$ for one user's top-$s$ topics set $\sigma$, and one topic $t$, which models a system similar to Warner's protocol for randomized response \cite{Warner1965}:\footnote{Instead of \qm{yes} or \qm{no} answers, we use $t \in \sigma$ or $t \notin \sigma$, respectively.}
\begin{align}
    \label{eq:theoretical-utility-counting-channel}
    \bordermatrix{
        \mathsf{C}_{\in} & t \in \sigma & t \notin \sigma \cr
        t \in \sigma & \mathtt{p} & 1 - \mathtt{p} \cr
        t \notin \sigma & \mathtt{q} & 1 - \mathtt{q} \cr
    }.
\end{align}

For $N$ Internet users, we consider the $N$-fold product of $\mathsf{C}_{\in}$ with itself post-processed by a counting query $T$, i.e. $\mathsf{C}_{\in}^{\otimes N} T$, defined as:
\begin{equation}
    \label{eq:counting-gain}
    T (\xi, \Xi) =
    \begin{cases}
        1 & \quad \text{if } \xi = \text{count}(\text{\qm{$t \in \sigma$}} \in \Xi) \text{,} \\
        0 & \quad \text{if } \xi \neq \text{count}(\text{\qm{$t \in \sigma$}} \in \Xi) \text{,}
    \end{cases}
\end{equation}
where $\Xi$ is the truthful list of values for the $N$ users, i.e. either \qm{$t \in \sigma$} or \qm{$t \notin \sigma$}, with $|\Xi| = N$, and $\xi$ is the analyst's noisy count of the number of occurrences of \qm{$t \in \sigma$}.
Hence, this gain function measures the probability of the analyst correctly counting the occurrences of topic $t$ in the population of Internet users.

Finally, we can compute the probability that the analyst can determine accurate total counts of topics by using Cor.~\ref{cor:kronecker-vulnerability} to evaluate the gain function $T$ with respect to the channel $\mathsf{C}_{\in}^{\otimes^N}$.

Defining $A = \nicefrac{\mathtt{p}}{\mathtt{q}}$, we compute the expected value of $A^{n}$ with respect to the binomial distribution, which gives us the probability of correctly counting the occurrences of a topic $t$ as a function of the Topics API parameters and of the population size $N$ of Internet users reporting topics, where $0 \leq n \leq N$:
\begin{equation}
    \label{eq:counting-binomial}
    \sum_{n=0}^{N} \binom{N}{n} \mathtt{p}^{n} (1-\mathtt{q})^{N-n} = \sum_{n=0}^{N} \binom{N}{n} A^{n} \mathtt{q}^{n} (1-\mathtt{q})^{N-n}.
\end{equation}

We compute this probability for the current values of the Topics API parameters in Sec.~\ref{sec:experimental-utility-counting}.

\subsubsection{A Gain Function for IBA Companies}
\label{sec:theoretical-utility-iba-gain}

Next, we would like to estimate the utility for IBA companies defined by the trustworthiness of the responses received. In particular, if an IBA company receives a topic $t$, what is the certainty with which a user indeed has an interest in $t$?
We propose a novel gain function for IBA companies using the Topics API, i.e. analysts that observe single reported topics.
This gain function measures whether a user's reported topic is genuine, i.e. from that user's top-$s$ set, or randomly chosen from the whole taxonomy.
Hence, the set of secrets contains top-$s$ sets, i.e. $\mathcal{X} = \Sigma$, where $\sigma \in \Sigma$ and $|\sigma| = s$, and the set of actions contains individual topics that may or may not be in a given top-$s$ set, i.e. $\mathcal{W} = \mathtt{T}$, where $t \in \mathtt{T}$ and $|\mathtt{T}| = m$.
Therefore, the IBA gain function $g_{\text{IBA}} : \mathtt{T} \times \Sigma \to \{ 0,1 \}$ is given by:
\begin{equation}
    \label{eq:iba-gain}
    g_{\text{IBA}} (t,\sigma) =
    \begin{cases}
        1 & \quad \text{if } t \in \sigma \text{,} \\
        0 & \quad \text{if } t \notin \sigma \text{.}
    \end{cases}
\end{equation}

We now derive the prior and posterior expected IBA gains and use them to compute this utility for Alg.\ref{alg:topics-report}.\footnote{For instance, when $(\pi_{\text{\{Music, News\}}}, \pi_{\text{\{Sports, Travel\}}}) = (\nicefrac{2}{3}, \nicefrac{1}{3})$, $N = 3$, $m' = 3$, $m = 4$, $s = 2$, and $r = 0.05$ in Fig.~\ref{fig:topics-channels}, $V_{g_{\text{IBA}}} (\pi) = 0.666$ and $V_{g_{\text{IBA}}} [\pi \triangleright \mathsf{C}_{\mathit{BN} \oplus_{r} \mathit{DP}}] = 0.977$. Considering a uniform prior instead, i.e. $(\pi_{\text{\{Music, News\}}}, \pi_{\text{\{Sports, Travel\}}}) = (\nicefrac{1}{2}, \nicefrac{1}{2})$, $V_{g_{\text{IBA}}} (\vartheta) = 0.5$ and $V_{g_{\text{IBA}}} [\vartheta \triangleright \mathsf{C}_{\mathit{BN} \oplus_{r} \mathit{DP}}] = 0.975$, which has a greater leakage.}

\begin{theorem}
    \label{theo:iba-gain-initial}
    The analyst prior expected gain considering the IBA gain function, $g_{\text{IBA}}$, and a prior probability distribution on top-$s$ sets, $\pi$, is given by the \emph{prior IBA vulnerability} (cf.~\eqref{eq:prior-vulnerability}):
    \begin{equation*}
        V_{g_{\text{IBA}}} (\pi) = \max_{t \in \mathtt{T}} \sum_{\sigma \in \Sigma \ : \ t \in \sigma} \pi_{\sigma}.
    \end{equation*}
\end{theorem}

Hence, the topic most likely to be genuine is the one with the maximum sum of the probabilities of all the top-$s$ sets it appears in.

\begin{theorem}
    \label{theo:iba-gain-final-bounded-noise}
    The analyst posterior expected gain considering the IBA gain function, $g_{\text{IBA}}$, the bounded noise channel, $\mathsf{C}_{\mathit{BN}}$, and a prior probability distribution on top-$s$ sets, $\pi$, is given by the \emph{posterior IBA vulnerability for bounded noise} (cf.~\eqref{eq:posterior-vulnerability}):
    \begin{equation*}
        V_{g_{\text{IBA}}} [\pi \triangleright \mathsf{C}_{\mathit{BN}}] = 1.
    \end{equation*}
\end{theorem}

Hence, the analyst can always distinguish a genuine topic from a random one up to the bounded noise step of the pipeline.

\begin{lemma}
    \label{lem:internal-choice-lower-bound}
    Given compatible channel matrices $\mathsf{C}$ and $\mathsf{D}$, i.e. both with the same input set $\mathcal{X}$, and their internal fixed-probability choice with probability $r$, $\mathsf{C} \ \oplus_{r} \mathsf{D}$, the posterior vulnerability of their internal probabilistic choice is bounded below by the maximum posterior vulnerability of the individual channels:
    \begin{equation*}
        V_{g} [\pi \triangleright \mathsf{C} \ \oplus_{r} \mathsf{D}] \geq \max \{ (1-r) \cdot V_{g} [\pi \triangleright \mathsf{C}], r \cdot V_{g} [\pi \triangleright \mathsf{D}] \}.
    \end{equation*}
\end{lemma}

\begin{theorem}
    \label{theo:iba-gain-final}
    The analyst posterior expected gain considering the IBA gain function, the final channel for utility analyses, $\mathsf{C}_{\mathit{BN} \oplus_{r} \mathit{DP}}$, and a prior probability distribution on top-$s$ sets, $\pi$, is given by the \emph{posterior IBA vulnerability for the Topics API} (cf.~\eqref{eq:posterior-vulnerability}):
    \begin{equation*}
        (1-r) \leq V_{g_{\text{IBA}}} [\pi \triangleright \mathsf{C}_{\mathit{BN} \oplus_{r} \mathit{DP}}] \leq (1-r) + r \cdot V_{g_{\text{IBA}}} (\pi).
    \end{equation*}
\end{theorem}

Hence, the analyst's posterior expected gain is bounded below by $(1-r)$, i.e. the probability that the Topics API reports a genuine topic, while above it can only reach $100\%$ if $r = 0$, i.e. if the Topics API has no differential privacy.

We consider scenarios for both channels $\mathsf{C}_{\mathit{BN}}$ and $\mathsf{C}_{\mathit{BN} \oplus_{r} \mathit{DP}}$.

\section{Experimental Evaluation}
\label{sec:experimental}

We start with a brief description of the datasets used on our experiments in Sec.~\ref{sec:experimental-datasets}, including ethical considerations in Sec.~\ref{sec:experimental-datasets-ethical}.
Then, we present the experimental results on our datasets for privacy vulnerability for individuals in Sec.~\ref{sec:experimental-privacy} and for utility for IBA companies in Sec.~\ref{sec:experimental-utility}, where we evaluate the vulnerability and the utility measures derived in Sec.~\ref{sec:theoretical-privacy} and Sec.~\ref{sec:theoretical-utility}, respectively.
We conclude in Sec.~\ref{sec:experimental-theoretical-limits} with a verification of the theoretical limits for the Topics API on our datasets under Google's current parameters.

\subsection{Datasets}
\label{sec:experimental-datasets}

\begin{table*}[t]
    \centering
    \begin{tabular}{|c|c|c|c|c|c|c|c|c|c|c|c|c|c|}
        \cline{2-14}
        \multicolumn{1}{c}{} & \multicolumn{4}{|c|}{Unique} & \multicolumn{9}{|c|}{Browsing history size} \\
        \hline
        \multicolumn{1}{|p{5.1em}|}{\centering AOL Dataset} & Users & URLs & Domains & Topics & Min. & 25\% & 50\% & 75\% & 95\% & 99\% & Max. & Mean & $\sigma$ \\
        \hline
        \hline
        \centering Original & 657\,426 & 1\,632\,789 & --- & --- & 1 & 5 & 17 & 52 & 228 & 566 & 279\,430 & 55.35 & 367.22 \\
        \hline
        \centering Experimental & 436\,005 & --- & 1\,291\,534 & --- & 2 & 5 & 14 & 42 & 181 & 439 & 6\,227 & 43.93 & 95.04 \\
        \hline
        \multicolumn{1}{|p{5.1em}|}{\centering Experimental \\ (Citizen Lab)} & 211\,313 & --- & 4\,872 & 31 & 2 & 3 & 6 & 13 & 47 & 116 & 2\,516 & 13.48 & 27.13 \\
        \hline
        \multicolumn{1}{|p{5.1em}|}{\centering Experimental \\ (Google v1)} & 198\,023 & --- & 2\,652 & 169 & 2 & 3 & 6 & 12 & 38 & 90 & 933 & 11.33 & 20.18 \\
        \hline
    \end{tabular}
    \caption{Experimental datasets statistics. The Original dataset is for reference only. Both datasets with topics classifications were derived from the Experimental dataset. All datasets span the period from 2006/03/01 to 2006/05/31, inclusive.}
    \label{tab:datasets-experiments}
\end{table*}

Our experiments were performed on treated versions of the AOL search logs dataset from $2006$ \cite{Pass2006}, which spans the period from $2006/03/01$ to $2006/05/31$, inclusive.
As proposed by Roesner et al. \cite{Roesner2012}, it may be used to reconstruct Internet users' browsing histories based on their visited websites.
Browsing history size statistics for the \emph{Original} and \emph{Experimental} datasets are presented in Tab.~\ref{tab:datasets-experiments}.

As described in Sec.~\ref{sec:models-cookies} and Sec.~\ref{sec:models-topics}, we assume adversaries and analysts capable of observing the whole browsing history and the whole top-$s$ topics set of every Internet user.
Hence, we have used the whole period of three months of available data as a single epoch for the Topics API.
We present in App.~\ref{app:model-longitudinal} how our model could be changed to account for more than one epoch.

\subsubsection{Ethical Considerations}
\label{sec:experimental-datasets-ethical}

The Original AOL search logs dataset was released in $2006$ under a \qm{non-commercial research use only} license.
However, it was shown that individuals on the dataset were vulnerable to re-identification via linkage attacks \cite{Barbaro2006}, which sparked ethical debates regarding its use \cite{Hafner2006,Anderson2006}.
Nevertheless, the dataset has been used in the past \cite{Bischoff2008,Heymann2008,Huang2009}, including for privacy analyses \cite{Roesner2012}, and it can still be found on the Internet \cite{WikipediaContributors2008,Arrington2006}.

Moreover, we have only used treated versions of the dataset on our experiments, as detailed in Sec.~\ref{sec:experimental-datasets-treated} and App.~\ref{app:datasets-treated}.
Our treated datasets have sizes (in number of rows) ranging from $6.17\%$ to $53.38\%$ of the Original dataset and we keep only each records' date and time, convert URLs to the respective domains, and convert every unique user identification number (\texttt{AnonID}) to a new random number (\texttt{RandID}) in a non-retrievable way.
Finally, we have been cleared by our Faculty Ethics Committee based on negligible risk.

\subsubsection{Experimental Dataset}
\label{sec:experimental-datasets-treated}

The first set of data treatments were performed to fix inconsistencies found on the Original AOL dataset, to drop unnecessary columns and rows without a URL, and to randomly remap every \texttt{AnonID} value to a new \texttt{RandID} value in a non-retrievable way as to avoid direct linkage of individuals from our treated datasets to the Original AOL dataset.\footnote{We have used Python's \href{https://docs.python.org/3/library/random.html\#random.SystemRandom}{\texttt{random.SystemRandom}} class for this task, which does not rely on software state and generates non-reproducible sequences.}

The second set of data treatments were performed to convert URLs to their respective domains.
This was necessary to guarantee realistic browsing histories for each user and to properly simulate the behavior of both third-party cookies and the current proposal of the Topics API, i.e. based on domains, not URLs.
To achieve that, we relied on the definition of \emph{effective top-level domain} (eTLD),\footnote{\url{https://developer.mozilla.org/en-US/docs/Glossary/eTLD}} i.e. \qm{a domain under which domains can be registered by a single organization}, on the technical requirements for domain names,\footnote{\qm{A domain name consists of one or more labels, each of which is formed from the set of ASCII letters, digits, and hyphens (a–z, A–Z, 0–9, -), but not starting or ending with a hyphen} \cite{WikipediaContributors2004,Berners-Lee1994,Fielding1995}.} and on an extended version of Mozilla's Public Suffix List.\footnote{\url{https://publicsuffix.org}}

The third set of data treatments were performed for consistency with the modeled scenarios and for better reproducibility of results.
We have dropped all users who have visited only one domain, including users who have only visited the same domain multiple times, and one outlier user.\footnote{The outlier user accounted for as many as $150\,802$ domain visits in an interval of three months. This accounts for $69$ domain visits per hour, $24$ hours a day, for $91$ days. Meanwhile, all other users have at most $6\,227$ domain visits in the same interval.}
We have also merged users' records by \texttt{RandID} to define their browsing histories as lists of tuples, with each tuple containing a visited domain and the respective date and time, and computed the top-$s$ sets of topics for each user.

The Experimental AOL dataset comprises $436\,005$ rows, one for each user (\texttt{RandID}), with the respective \texttt{BrowsingHistory} list.
This accounts for $19\,154\,979$ records, i.e. individual domain visits, and for $1\,291\,534$ unique domains.

The following sets of data treatments were performed to assign topics to domains and to remove domains without a classification.
We have used two distinct classifications:\footnote{We are not evaluating machine classification models, so we have opted to use only human-based, static classifications. See \cite{Beugin2023} for empirical analyses of the sort.} Citizen Lab's URL testing lists for Internet censorship \cite{CitizenLab2014}, presented in Sec.~\ref{sec:experimental-datasets-citizen-lab}, and Google's static classification of domains provided with the Chrome browser, presented in Sec.~\ref{sec:experimental-datasets-google}.

\subsubsection{Citizen Lab Classification}
\label{sec:experimental-datasets-citizen-lab}

The original Citizen Lab URL testing lists dataset accounts for $33\,861$ unique URLs and $31$ distinct categories.\footnote{\href{https://github.com/citizenlab/test-lists/tree/ebd0ee8d41977b381972b2f6c471af5437d8d015/lists}{https://github.com/citizenlab/test-lists}, commit \texttt{ebd0ee8}, merge of all lists and using the new category codes.}
We have performed the same data treatment described in Sec.~\ref{sec:experimental-datasets-treated} to convert URLs to their respective domains, resulting in $27\,792$ rows, one for each \texttt{domain} and their set of \texttt{topics}, further reduced once matched with domains on the treated AOL dataset.

The Experimental AOL dataset with Citizen Lab's classification, after dropping all users who have visited only one domain, including users who have only visited the same domain multiple times, and the same outlier user, comprises $211\,313$ rows, one for each user (\texttt{RandID}), with the respective \texttt{BrowsingHistory} and \texttt{sTopics} lists.
This accounts for $2\,847\,578$ records, i.e. individual domain visits, and for $4\,872$ unique domains.

\subsubsection{Google Topics Classification}
\label{sec:experimental-datasets-google}

The static domains classification provided by Google with the Chrome browser, after extraction, comprises $9\,046$ rows, one for each \texttt{domain} and the respective set of \texttt{topics} by code, according to Google's Topics taxonomy v1, which accounts for $349$ distinct categories.\footnote{\url{https://github.com/patcg-individual-drafts/topics/blob/main/taxonomy_v1.md}}
We have treated this dataset only to remove domains without a classification, i.e. $1\,315$ rows.
The number of rows is further reduced once matched with domains on the treated AOL dataset.

The Experimental AOL dataset with Google's Topics classification, after dropping all users who have visited only one domain, including users who have only visited the same domain multiple times, and the same outlier user, comprises $198\,023$ rows, one for each user (\texttt{RandID}), with the respective \texttt{BrowsingHistory} and \texttt{sTopics} lists.
This accounts for $2\,243\,660$ records, i.e. individual domain visits, and for $2\,652$ unique domains.

We summarize in Tab.~\ref{tab:datasets-experiments} some statistics of the Original and Experimental AOL datasets.
More detailed steps of the data treatments performed are provided in App.~\ref{app:datasets}.

\subsection{Privacy Vulnerability for Individuals}
\label{sec:experimental-privacy}

\begin{table}[t]
    \centering
    \begin{tabular}{|c|c|c|c|c|}
        \cline{2-5}
        \multicolumn{1}{c}{} & \multicolumn{4}{|c|}{Privacy: (average-case) channel capacity} \\
        \cline{2-5}
        \multicolumn{1}{}{} & \multicolumn{1}{|c|}{Cookies} & \multicolumn{3}{|c|}{Topics API} \\
        \hline
        AOL Dataset & $\mathsf{C}_{\mathcal{C}}$ & $\mathsf{C}_{\mathit{BH}} \mathsf{C}_{G}$ & $\mathsf{C}_{\mathit{BH}} \mathsf{C}_{G} \mathsf{C}_{\mathit{BN}}$ & $\mathsf{C}_{\mathcal{T}}$ \\
        \hline
        \hline
        \centering Experimental & $436\,005$ & --- & --- & --- \\
        \hline
        \multicolumn{1}{|p{5.1em}|}{\centering Experimental \\ (Citizen Lab)} & $211\,313$ & $27\,970$ & $6.20$ & $5.94$ \\
        \hline
        \multicolumn{1}{|p{5.1em}|}{\centering Experimental \\ (Google v1)} & $198\,023$ & $110\,870$ & $33.80$ & $32.16$ \\
        \hline
    \end{tabular}
    \caption{Privacy results for (average-case) channel capacities, i.e. the multiplicative Bayes leakage under a uniform prior probability distribution on Internet users, for a single epoch, where $\mathsf{C}_{\mathcal{C}} = \mathsf{C}_{\mathit{BH}} \mathsf{C}_{U}$, $\mathsf{C}_{\mathcal{T}} = \mathsf{C}_{\mathit{BH}} \mathsf{C}_{G} \mathsf{C}_{\mathit{BN} \oplus_{0.05} \mathit{DP}}$, $s=5$, $r=0.05$.}
    \label{tab:results-privacy-bayes}
\end{table}

We present in Tab.~\ref{tab:results-privacy-bayes} the (average-case) channel capacities for all the considered scenarios for third-party cookies and the Topics API, using the current parameters set by Google, i.e. $s=5$ and $r=0.05$.

For third-party cookies, i.e. $\mathsf{C}_{\mathcal{C}}$, we observe values of leakage equal to the respective number of Internet users on each of the Experimental datasets.
As expected, an adversary using third-party cookies to track individuals' browsing histories would always succeed on average, given access to auxiliary information to match names to browsing histories and their unique identifiers.

For the Topics API, we consider three incremental scenarios: generalization only, i.e. $\mathsf{C}_{\mathit{BH}} \mathsf{C}_{G}$; generalization and bounded noise, i.e. $\mathsf{C}_{\mathit{BH}} \mathsf{C}_{G} \mathsf{C}_{\mathit{BN}}$; and generalization and bounded noise with differential privacy, i.e. $\mathsf{C}_{\mathcal{T}}$, the complete pipeline.
Generalization alone is able to reduce leakage to $13.24\%$ (Citizen Lab) and to $55.99\%$ (Google v1) of the respective leakages for third-party cookies, results that are clearly dependent on the sizes of the respective classification taxonomies.
The addition of bounded noise, which also changes the output of the channel from top-$s$ topics sets to individual topics, further reduces leakage to $0.0029\%$ (Citizen Lab) and to $0.0171\%$ (Google v1) of the respective leakages for third-party cookies.
Finally, the addition of differential privacy on $5\%$ of the reported topics slightly decreases leakage to $0.0028\%$ (Citizen Lab) and to $0.0162\%$ (Google v1) of the respective leakages for third-party cookies, but also introduces plausible deniability to the Topics API.

The leakage values for third-party cookies and for the Topics API up to the Generalization step differ among the datasets in Tab.~\ref{tab:results-privacy-bayes} due to the different number of columns on each dataset.
In particular, the channel up to Generalization, $\mathsf{C}_{\mathit{BH}} \mathsf{C}_{G}$, does not reach capacity for any dataset.\footnote{Citizen Lab: $27\,970$ (leakage, i.e. number of columns) $<$ $169\,911$ (capacity, i.e. all combinations). Google v1: $110\,870$ (leakage) $<$ $1\,082\,239\,158$ (capacity).}
For the Topics API up to the Bounded Noise step and up to the Differential Privacy step, the leakage values differ among the datasets due to the parameter $m$, since in both cases the respective channels, $\mathsf{C}_{\mathit{BH}} \mathsf{C}_{G} \mathsf{C}_{\mathit{BN}}$ and $\mathsf{C}_{\mathcal{T}}$, reach capacity.

\subsection{Utility for IBA Companies}
\label{sec:experimental-utility}

\subsubsection{Learning the Most Popular Topic}
\label{sec:experimental-utility-counting}

\begin{figure}[t]
    \centering
    \includegraphics[width=0.45\textwidth]{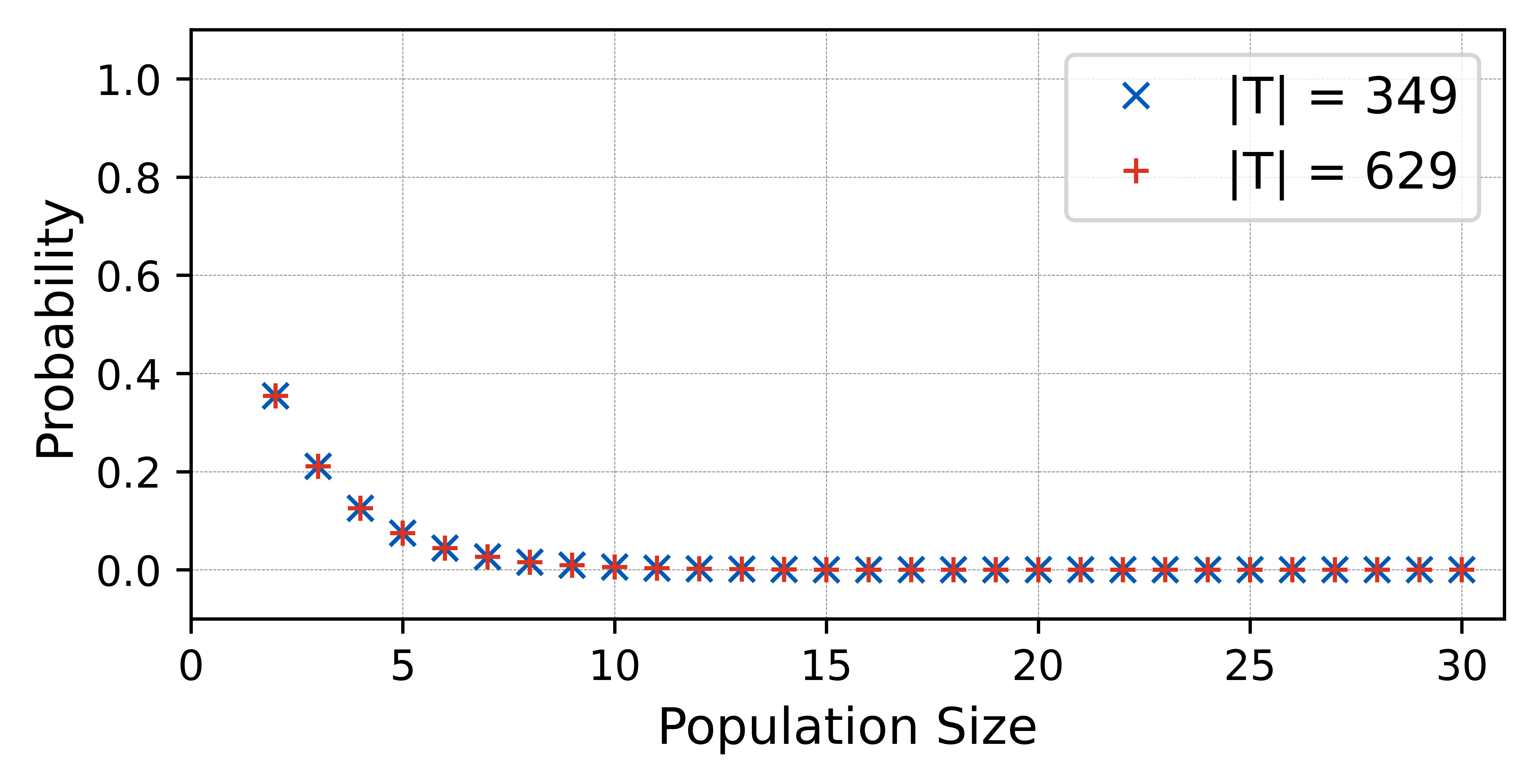}
    \caption{Expected value for $A^{n}$ with respect to the binomial distribution (cf.~\eqref{eq:counting-binomial}), where $A = \nicefrac{\mathtt{p}}{\mathtt{q}}$, $\mathtt{p} = \nicefrac{1-r}{s} + \nicefrac{r}{m}$, and $\mathtt{q} = \nicefrac{r}{m}$, according to channel $\mathsf{C}_{\mathit{BN} \oplus_{r} \mathit{DP}}$ (cf.~\eqref{eq:model-topics-utility-channel}), for $2 \leq N \leq 30$, $s=5$, $m=349$ or $m=629$, and $r=0.05$. As the population size of Internet users reporting topics increases, the probability of correctly counting the occurrences of a topic $t$ in all the top-$s$ sets approaches zero for both taxonomy sizes.}
    \Description{A plot with Probability on the vertical axis, with values 0.0, 0.2, 0.4, 0.6, 0.8, and 1.0, and with Population Size on the horizontal axis, with values 0, 5, 10, 15, 20, 25, and 30. Blue multiplication symbols represent data for a taxonomy of size 349, and red summation symbols represent data for a taxonomy of size 629. For every Population Size, both blue and red symbols seem to overlap exactly with the same Probability. For Population Sizes of 2, 3, 4, and 5, the Probability is just above 0.35, just above 0.20, just below 0.15, and just below 0.10, respectively. For a Population Size of 10, the Probability reaches close to 0.00, and continues close to 0.00 until a Population Size of 30.}
    \label{fig:counting}
\end{figure}

Considering the current values for the Topics API parameters, $s=5$, $m=349$ or $m=629$, and $r=0.05$,\footnote{Which results in $A=1327.2$ and $\mathtt{q}=0.00014326647564469913$ for $m=349$, and in $A=2391.2$ and $\mathtt{q}=0.00007949125596184421$ for $m=629$.} we show in Fig.~\ref{fig:counting}, for population sizes of Internet users reporting topics between $2 \leq N \leq 30$, that noise quickly builds-up for the counting query $T$ (cf.~\eqref{eq:counting-gain}).
For instance, for a population of only $10$ individuals, the chance of an analyst correctly counting the occurrence of a topic $t$ is as low as $0.56\%$ for both taxonomy sizes.

\subsubsection{IBA expected gain}
\label{sec:experimental-utility-iba-gain}

\begin{table}[t]
    \centering
    \begin{tabular}{|c|c|c|c|c|}
        \cline{2-5}
        \multicolumn{1}{c}{} & \multicolumn{4}{|c|}{Utility: Bayes vulnerability} \\
        \cline{2-5}
        \multicolumn{1}{c}{} & \multicolumn{1}{|c|}{Cookies} & \multicolumn{3}{|c|}{Topics API} \\
        \cline{2-5}
        \multicolumn{1}{c}{} & \multicolumn{1}{|c|}{$\mathsf{C}_{U}$} & \multicolumn{1}{|c|}{$\mathsf{C}_{G}$} & \multicolumn{1}{|c|}{$\mathsf{C}_{\mathit{BN}}$} & \multicolumn{1}{|c|}{$\mathsf{C}_{\mathit{BN} \oplus_{0.05} \mathit{DP}}$} \\
        \hline
        AOL Dataset & Posterior & Posterior & Posterior & Posterior \\
        \hline
        \hline
        \multicolumn{1}{|p{5.1em}|}{\centering Experimental} & $1.0$ & --- & --- & --- \\
        \hline
        \multicolumn{1}{|p{5.1em}|}{\centering Experimental \\ (Citizen Lab)} & $1.0$ & $0.132$ & $0.107$ & $0.102$ \\
        \hline
        \multicolumn{1}{|p{5.1em}|}{\centering Experimental \\ (Google v1)} & $1.0$ & $0.560$ & $0.048$ & $0.046$ \\
        \hline
    \end{tabular}
    \caption{Utility results for Bayes vulnerability under uniform prior probability distributions on browsing histories, for channels $\mathsf{C}_{U}$ and $\mathsf{C}_{G}$, or under non-uniform prior probability distributions on top-$s$ sets, for channels $\mathsf{C}_{\mathit{BN}}$ or $\mathsf{C}_{\mathit{BN} \oplus_{0.05} \mathit{DP}}$, for a single epoch. The prior Bayes vulnerabilities for the Experimental dataset and $\mathsf{C}_{U}$ channel is $2.3 \cdot 10^{-6}$; for the Experimental (Citizen Lab) dataset and both $\mathsf{C}_{U}$ and $\mathsf{C}_{G}$ channels is $4.7 \cdot 10^{-6}$; for the Experimental (Citizen Lab) dataset and both $\mathsf{C}_{\mathit{BN}}$ or $\mathsf{C}_{\mathit{BN} \oplus_{0.05} \mathit{DP}}$ channels is $0.032$; for the Experimental (Google v1) dataset and both $\mathsf{C}_{U}$ and $\mathsf{C}_{G}$ channels is $5.0 \cdot 10^{-6}$; and for the Experimental (Google v1) dataset and both $\mathsf{C}_{\mathit{BN}}$ or $\mathsf{C}_{\mathit{BN} \oplus_{0.05} \mathit{DP}}$ channels is $0.015$.}
    \label{tab:results-utility-bayes}
\end{table}

\begin{table}[t]
    \centering
    \begin{tabular}{|c|c|c|}
        \cline{2-3}
        \multicolumn{1}{c}{} & \multicolumn{2}{|c|}{Utility: IBA expected gain} \\
        \cline{2-3}
        \multicolumn{1}{c}{} & \multicolumn{2}{|c|}{Topics API} \\
        \cline{2-3}
        \multicolumn{1}{c}{} & \multicolumn{1}{|c|}{$\mathsf{C}_{\mathit{BN}}$} & \multicolumn{1}{|c|}{$\mathsf{C}_{\mathit{BN} \oplus_{0.05} \mathit{DP}}$} \\
        \hline
        AOL Dataset & Posterior & Posterior \\
        \hline
        \hline
        \multicolumn{1}{|p{5.1em}|}{\centering Experimental \\ (Citizen Lab)} & $1.000$ & $0.958$ \\
        \hline
        \multicolumn{1}{|p{5.1em}|}{\centering Experimental \\ (Google v1)} & $1.000$ & $0.962$ \\
        \hline
    \end{tabular}
    \caption{Utility results for IBA expected gain under non-uniform prior probability distributions on top-$s$ topics sets for a single epoch. The prior IBA expected gain for the Experimental (Citizen Lab) dataset is $0.720$, and for the Experimental (Google v1) dataset is $0.421$. The posterior IBA expected gain for both datasets is above $0.95$, in agreement with Thm.~\ref{theo:iba-gain-final}.}
    \label{tab:results-utility-iba}
\end{table}

We present in Tab.~\ref{tab:results-utility-bayes} the (average-case) Bayes vulnerabilities and in Tab.~\ref{tab:results-utility-iba} the IBA expected gain, i.e. the IBA vulnerabilities, for all the considered scenarios for both third-party cookies and the Topics API.
We report vulnerabilities instead of leakages because we are not always considering uniform prior probability distributions, but also non-uniform, data-dependent prior probability distributions.
Specifically, the prior probability distributions on browsing histories, i.e. for $\mathsf{C}_{U}$ and $\mathsf{C}_{G}$, are uniform, while those on top-$s$ topics sets, i.e. for $\mathsf{C}_{\mathit{BN}}$ and $\mathsf{C}_{\mathit{BN} \oplus_{0.05} \mathit{DP}}$ are non-uniform and data-dependent.
The multiplicative leakages can be computed from the reported values according to Eq.~\ref{eq:leakage}.

First, we analyze the results for Bayes vulnerability in Tab.~\ref{tab:results-utility-bayes}, i.e. how likely the analyst is to completely reconstruct browsing histories or top-$s$ sets on the first try.
(Notice that the leakage results for $\mathsf{C}_{U}$ and $\mathsf{C}_{G}$ are also their respective channel capacities.)
For third-party cookies, i.e. $\mathsf{C}_{U}$, the posterior vulnerabilities equal $100\%$ for all the Experimental datasets.
As expected, an analyst using third-party cookies to gather browsing history data would always succeed on average.

For the Topics API, we consider three incremental scenarios: generalization only, i.e. $\mathsf{C}_{G}$; generalization and bounded noise, i.e. $\mathsf{C}_{\mathit{BN}}$; and generalization and bounded noise with differential privacy, i.e. $\mathsf{C}_{\mathit{BN} \oplus_{0.05} \mathit{DP}}$.
Generalization alone is able to reduce the probability of an analyst correctly reconstructing browsing histories from top-$s$ sets to $13.2\%$ (Citizen Lab) and $56.0\%$ (Google v1), a clear dependence on the sizes of the respective classification taxonomies.
The addition of bounded noise, which also changes the input of the channel from browsing histories to top-$s$ sets and the output from top-$s$ sets to individual topics, further reduces the probability of an analyst correctly reconstructing the top-$s$ sets from observed topics to $10.7\%$ (Citizen Lab) and $4.8\%$ (Google v1).
Finally, the addition of differential privacy on $5\%$ of the reported topics slightly decreases the probabilities to $10.2\%$ (Citizen Lab) and $4.6\%$ (Google v1).

Second, we analyze the results for the IBA expected gain, i.e. the IBA vulnerabilities, in Tab.~\ref{tab:results-utility-iba}.
Notice that the IBA gain function is only defined for channels $\mathsf{C}_{\mathit{BN}}$ and $\mathsf{C}_{\mathit{BN} \oplus_{r} \mathit{DP}}$, and that both channels have as inputs top-$s$ sets and as outputs individual topics.
As expected, bounded noise alone would allow an analyst to always correctly guess if an observed topic comes from a top-$s$ set since there is no randomness on topics reported.
The addition of differential privacy on $5\%$ of the reported topics slightly decreases the analyst's certainty to $95.8\%$ (Citizen Lab) and to $96.2\%$ (Google v1).

\subsection{Theoretical Limits for the Topics API}
\label{sec:experimental-theoretical-limits}

\begin{table}[t]
    \centering
    \begin{tabular}{|c|c|c|c|c|}
        \cline{2-5}
        \multicolumn{1}{c}{} & \multicolumn{4}{|c|}{Privacy: theoretical limits} \\
        \hline
        \multicolumn{1}{|p{5.1em}|}{\centering AOL Dataset} & $m$ & $\mathcal{ML}_{1}^{\times}$ & $\epsilon_{\mathsf{C}_{\mathcal{T}}}$ & $e^{\epsilon_{\mathsf{C}_{\mathcal{T}}}}$ \\
        \hline
        \hline
        \multicolumn{1}{|p{5.1em}|}{\centering Experimental \\ (Citizen Lab)} & $31$ & $5.94$ & $4.777$ & $118.8$ \\
        \hline
        \multicolumn{1}{|p{5.1em}|}{\centering Experimental \\ (Google v1)} & $169$ & $32.16$ & $6.466$ & $643.2$ \\
        \hline
    \end{tabular}
    \caption{Theoretical limits for the Experimental datasets with topics classification, considering $s=5$ and $r=0.05$, for: the multiplicative (average-case) Bayes capacity, $\mathcal{ML}_{1}^{\times}$ (cf.~\eqref{eq:theoretical-privacy-topics-average-complete-leakage}); the differential privacy measure of indistinguishability for the discrete metric, $\epsilon_{\mathsf{C}_{\mathcal{T}}}$ (cf.~\eqref{eq:theoretical-privacy-topics-epsilon}); and the (max-case) capacity, $e^{\epsilon_{\mathsf{C}_{\mathcal{T}}}} = \mathcal{ML}_{g}^{\max} (\mathsf{C}_{\mathcal{T}})$ (cf.~\eqref{eq:theoretical-privacy-topics-max-complete-capacity}).}
    \label{tab:theoretical-aol}
\end{table}

We present in Tab.~\ref{tab:theoretical-aol} the theoretical values for the multiplicative (average-case) Bayes capacity, $\mathcal{ML}_{1}^{\times}$ (cf.~\eqref{eq:theoretical-privacy-topics-average-complete-leakage}), the differential privacy measure of indistinguishability for the discrete metric, $\epsilon_{\mathsf{C}_{\mathcal{T}}}$ (cf.~\eqref{eq:theoretical-privacy-topics-epsilon}), and the (max-case) capacity $e^{\epsilon_{\mathsf{C}_{\mathcal{T}}}} = \mathcal{ML}_{g}^{\max} (\mathsf{C}_{\mathcal{T}})$ (cf.~\eqref{eq:theoretical-privacy-topics-max-complete-capacity}), for the Experimental datasets with topics classifications.
We consider the size of the top-$s$ sets as $s=5$, the probability of reporting a random topic from the whole taxonomy as $r = 0.05$, and the size of each taxonomy as the total number of distinct topics that occur on each dataset (as in Tab.~\ref{tab:datasets-experiments}).

\subsubsection{Privacy Vulnerability for Individuals}
\label{sec:experimental-theoretical-limits-privacy}

The Bayes capacities in Tab.~\ref{tab:theoretical-aol}, i.e. $5.94$ (Citizen Lab) and $32.16$ (Google v1), validate our results on the last column of Tab.~\ref{tab:results-privacy-bayes} for the channel $\mathsf{C}_{\mathcal{T}}$.
Moreover, we have in Tab.~\ref{tab:theoretical-aol} the differential privacy measure of indistinguishability for the discrete metric, $\epsilon_{\mathsf{C}_{\mathcal{T}}}$, i.e. $4.777$ (Citizen Lab) and $6.466$ (Google v1), which allows us to compute the (max-case) capacity for the channel $\mathsf{C}_{\mathcal{T}}$, $e^{\epsilon_{\mathsf{C}_{\mathcal{T}}}} = \mathcal{ML}_{g}^{\max} (\mathsf{C}_{\mathcal{T}})$, i.e. $118.8$ (Citizen Lab) and $643.2$ (Google v1), which captures worst-case scenarios of privacy vulnerability for individuals.
Hence, (average-case) leakage reaches capacity for both of our datasets, while the worst-case capacity means some individuals have their probability of re-identification increased up to $643$ times.

\subsubsection{Utility for IBA Companies}
\label{sec:experimental-theoretical-limits-utility}

The multiplicative (average-case) Bayes leakages over non-uniform prior distributions computed from Tab.~\ref{tab:results-utility-bayes} for the channel $\mathsf{C}_{\mathit{BN} \oplus_{0.05} \mathit{DP}}$ equal $3.18$ (Citizen Lab) and $3.06$ (Google v1), and the multiplicative leakages for the IBA gain function for the same channel computed from Tab.~\ref{tab:results-utility-iba} equal $1.33$ (Citizen Lab) and $2.28$ (Google v1).
All the results are within the respective capacities, $5.94$ (Citizen Lab) and $32.16$ (Google v1).
Hence, analysts are poorly equipped to reconstruct all top-$s$ sets from observed topics and, as discussed in Sec.~\ref{sec:experimental-utility-counting}, to draw statistical conclusions on a topic popularity, while being highly confident of whether a topic is genuine or not with high posteriors in Tab.~\ref{tab:results-utility-iba}.

\section{Discussion}
\label{sec:discussion}

\begin{table}[t]
    \centering
    \begin{tabular}{|c|c|c|c|c|}
        \cline{2-5}
        \multicolumn{1}{c}{} & \multicolumn{4}{|c|}{Privacy: theoretical limits} \\
        \cline{2-5}
        \hline
        \multicolumn{1}{|p{6.5em}|}{\centering Scenario} & $m$ & $\mathcal{ML}_{1}^{\times}$ & $\epsilon_{\mathsf{C}_{\mathcal{T}}}$ & $e^{\epsilon_{\mathsf{C}_{\mathcal{T}}}}$ \\
        \hline
        \hline
        \multicolumn{1}{|p{6.5em}|}{\centering Google Topics \\ API v1} & $349$ & $66.36$ & $7.191$ & $1\,327.2$ \\
        \hline
        \multicolumn{1}{|p{6.5em}|}{\centering Google Topics \\ API v2} & $629$ & $119.56$ & $7.780$ & $2\,391.2$ \\
        \hline
        \multicolumn{1}{|p{6.5em}|}{\centering Google Natural \\ Language v2} & $1\,091$ & $207.34$ & $8.330$ & $4\,146.8$ \\
        \hline
        \multicolumn{1}{|p{6.5em}|}{\centering IAB Audience \\ Taxonomy v1.1} & $1\,679$ & $319.06$ & $8.761$ & $6\,381.2$ \\
        \hline
    \end{tabular}
    \caption{Theoretical limits for the Google Topics API proposed taxonomies and related ones, considering $s=5$ and $r=0.05$, for: the multiplicative (average-case) Bayes capacity, $\mathcal{ML}_{1}^{\times}$ (cf.~\eqref{eq:theoretical-privacy-topics-average-complete-leakage}); the differential privacy measure of indistinguishability for the discrete metric, $\epsilon_{\mathsf{C}_{\mathcal{T}}}$ (cf.~\eqref{eq:theoretical-privacy-topics-epsilon}); and the (max-case) capacity $e^{\epsilon_{\mathsf{C}_{\mathcal{T}}}} = \mathcal{ML}_{g}^{\max} (\mathsf{C}_{\mathcal{T}})$ (cf.~\eqref{eq:theoretical-privacy-topics-max-complete-capacity}).}
    \label{tab:theoretical-google}
\end{table}

Our results from Sec.~\ref{sec:experimental} for the whole Topics API pipeline, i.e. channel $\mathsf{C}_{\mathcal{T}}$, show that the privacy vulnerability for worst-case scenarios are at least one order of magnitude greater than their equivalent average-case scenarios.
Moreover, a greater than five-fold increase on the average-case channel capacity is only accompanied by a $35\%$ increase on the value of the indistinguishability parameter $\epsilon$.
The choice of values for the parameter $\epsilon$ is usually regarded as a \qm{social question} \cite{Dwork2008} and values ranging from $0.01$ to $10$ are usually considered \qm{safe} but without solid grounds for those choices \cite{Hsu2014}.
Our results highlight that relying only on the value of $\epsilon$ for privacy analyses may be misleading.

We present in Tab.~\ref{tab:theoretical-google} the theoretical limits for the taxonomies proposed by Google for the Topics API, Google Topics API v1 ($349$ categories\footnote{\url{https://github.com/patcg-individual-drafts/topics/blob/main/taxonomy_v1.md}}) and Google Topics API v2 ($629$ categories\footnote{\url{https://github.com/patcg-individual-drafts/topics/blob/main/taxonomy_v2.md}}), and two additional, related taxonomies, Google Natural Language v2 ($1\,091$ categories\footnote{\url{https://www.google.com/basepages/producttype/taxonomy-with-ids.en-US.txt}}) and IAB Audience Taxonomy v1.1 ($1\,679$ categories\footnote{\url{https://iabtechlab.com/standards/audience-taxonomy}}), considering $s = 5$ and $r = 0.05$.
Given that the channel capacities for both average- (cf.~\eqref{eq:theoretical-privacy-topics-average-complete-leakage}) and worst-case (cf.~\eqref{eq:theoretical-privacy-topics-max-complete-capacity}) scenarios are directly proportional to the taxonomy size, the already implemented taxonomy change from the Google Topics API v1 to v2 has increased the channel capacity in $80.17\%$.
The adoption of the Google Natural Language v2 would increase it in $212.45\%$, while IAB Audience Taxonomy v1.1 would increase it in $380.80\%$, both with respect to the Google Topics API v1 taxonomy.

Hence, changes to the taxonomy size ($m$) should be balanced by changes on the other two parameters of the Topics API, i.e. the top-$s$ set size ($s$) and the probability of returning a random topic from the whole taxonomy instead of a random topic from a user's top-$s$ set ($r$).
However, $r$ directly impacts the lower bound of the equation on Thm.~\ref{theo:iba-gain-final}, i.e. the API's utility measured as the trustworthiness of the reported topic, so changes on $r$ would be undesirable for IBA.

For instance, suppose the channel capacities for the Google Topics API v1 taxonomy ($m=349$) with $r=0.05$ and $s=5$, i.e. $66.36$ (average-case) and $1\,327.20$ (max-case), were agreed to be acceptable and increases should be avoided.
In order to increase the taxonomy size to implement the Google Topics API v2 taxonomy ($m=629$) while keeping $r=0.05$ and similar channel capacities as for $m=349$, it would be necessary to have $s=9$.
In summary, keeping $r=0.05$ and channel average- and max-case capacities mostly unchanged, increases on the taxonomy size ($m$) should be accompanied by proportional increases on the top-$s$ set size ($s$).

\begin{table}[t]
    \centering
    \begin{tabular}{|c|c|c|c|c|}
        \hline
        \multicolumn{5}{|c|}{Privacy: theoretical limits} \\
        \hline
        $m$ & $r$ & $s$ & $\mathcal{ML}_{1}^{\times}$ & $\mathcal{ML}_{g}^{\max}$ \\
        \hline
        \hline
        $349$ & $0.05$ & $5$ & $66.36$ & $1\,327.20$ \\
        \hline
        $629$ & $0.05$ & $5$ & $119.56$ & $2\,391.20$ \\
        \hline
        $629$ & $0.47$ & $5$ & $67.14$ & $142.86$ \\
        \hline
        $629$ & $0.37$ & $6$ & $66.42$ & $179.50$ \\
        \hline
        $629$ & $0.26$ & $7$ & $66.75$ & $256.75$ \\
        \hline
        $629$ & $0.15$ & $8$ & $66.98$ & $446.54$ \\
        \hline
        $629$ & $0.05$ & $9$ & $66.44$ & $1\,328.89$ \\
        \hline
    \end{tabular}
    \caption{The Topics API multiplicative (average-case) Bayes capacity, $\mathcal{ML}_{1}^{\times}$ (cf.~\eqref{eq:theoretical-privacy-topics-average-complete-leakage}), and (max-case) capacity, $\mathcal{ML}_{g}^{\max}$ (cf.~\eqref{eq:theoretical-privacy-topics-max-complete-capacity}), for different values of the parameters $m$, $s$, and $r$.}
    \label{tab:theoretical-worst-case}
\end{table}

Interestingly, increases on the value of $r$, even though unlikely, would have a considerable positive impact on the channel worst-case capacity, and hence decrease the likelihood of re-identification for outliers and worst-case privacy scenarios, as presented in Tab.~\ref{tab:theoretical-worst-case}.

Finally, we could have broken the channel $\mathsf{C}_{G}$ (lines 1--3 of Alg.~\ref{alg:topics-derive}) to account for two additional steps in the Topics API pipeline apart from generalization and that might include randomness: the addition of random topics to top-$s$ sets that have less than $s$ topics, and the pruning of sets with more than $s$ topics.
Those are data-dependent steps that would still produce results within our reported bounds, e.g. $r + \frac{m' (1 - r)}{s} = \mathcal{L}_{1}^{\times} (\vartheta, \mathsf{C}_{\mathcal{T}}) \leq r + \frac{m (1 - r)}{s}$ in Eq.~\ref{eq:theoretical-privacy-topics-average-complete-leakage}.

\section{Related Work}
\label{sec:related-work}

This work is a substantial extension of our previous, preliminary study on privacy in the Topics API \cite{Alvim2023a}.

Google published two theoretical and empirical privacy analyses of the Topics API \cite{Epasto2022,Carey2023}, including an estimated $3\%$ chance of correct re-identification of a random Internet user, but their analyses do not account for worst-case scenarios, as observed by Mozilla's Martin Thomson \cite{Thomson2023}.
Moreover, Google's upper-bound results are not generally valid and depend on \qm{closed-world} assumptions, i.e. adversaries without access to side information, and their model does not provide a value for the differential privacy parameter $\epsilon$.

Beugin et al. \cite{Beugin2023}\footnote{They also empirically evaluated the API classification of contexts as a utility measure.} empirically demonstrated that third-parties are able to distinguish between real and random topics, and attributed this result to the stability of Internet users' interests across epochs.
This result is in accordance with our experimental results for the IBA expected gain, presented in Tab.~\ref{tab:results-utility-iba}, which corroborates the lower bound given in Thm.~\ref{theo:iba-gain-final}.

Jha et al. \cite{Jha2023} also empirically demonstrated the inefficacy of reporting random topics and that re-identification of Internet users is possible if third-parties observe users across enough epochs.\footnote{We show how to extend our model to account for more than one epoch in App.~\ref{app:model-longitudinal}.}
This result is in accordance with our theoretical limits (Tab.~\ref{tab:theoretical-aol}) and experimental results (Tab.~\ref{tab:results-privacy-bayes}), which show that the (average-case) leakage reaches capacity on both of our datasets for the Topics API.
This capacity represents a much lower, but non-zero probability of re-identification if compared to third-party cookies.

Roesner et al. \cite{Roesner2012} also used the AOL search logs dataset \cite{Pass2006} to reconstruct Internet users' browsing histories based on their visited URLs.
Olejnik et al. \cite{Olejnik2014} did similar analyses of Internet users' browsing histories and derived category profiles (i.e. topics of interest, but not limited to a fixed number per user).
They empirically demonstrated that $88\%$ of category profiles could be attributed to a unique user, and that taxonomies of only $30$ categories were enough to replicate the \qm{long tail pattern} seen on browsing histories.
The same study was replicated by Bird et al. \cite{Bird2020} with similar results.

Quantitative Information Flow was pioneered by Clark, Hunt, and Malacaria \cite{Clark2002}, followed by a growing community, e.g. \cite{Malacaria2007,Chatzikokolakis2008,Smith2009}, and its principles have been organized in~\cite{Alvim2020}. Moreover, the Haskell-based Kuifje programming language \cite{Gibbons2020,Bognar2019} is able to interpret exactly the code given in the algorithms provided here.

\section{Conclusion}
\label{sec:conclusion}

We presented a novel model for the Topics API including theoretical and experimental results for the privacy vulnerability for individuals and for the utility for IBA companies.
Our theoretical results for privacy vulnerability include upper-bounds for average- and worst-case scenarios, and account for unknown, arbitrary correlations that could be used by an adversary as side information.

\paragraph{Availability}
Our privacy and utility analyses \cite{Nunes2024a}, together with the Treated, Reduced, and Experimental AOL datasets \cite{Nunes2024b}, are available on Zenodo under \href{https://choosealicense.com/licenses/gpl-3.0/}{GNU GPLv3} and \href{https://creativecommons.org/licenses/by-nc-sa/4.0/}{CC BY-NC-SA 4.0} licenses, respectively.

\begin{acks}
Mário S.\ Alvim and Gabriel H.\ Nunes were supported by CNPq, CAPES, and FAPEMIG.
The authors are grateful to Andrés Muñoz Medina from Google Research for the constructive interactions.
\end{acks}

\balance
\printbibliography

\newpage
\appendix

\section{Proofs}
\label{app:proofs}

\allowdisplaybreaks

We start by introducing an additional QIF definition.

\paragraph{External Fixed-Probability Choice}
This composition of two compatible channel matrices, i.e. both with the same input set $\mathcal{X}$, models a scenario in which the system can (publicly) choose between two possible paths according to a fixed-probability $r$.
It differs from the internal fixed-probability choice, Eq.~\ref{eq:internal-choice}, because the chosen path is known to observers of the system.

Formally, given compatible channel matrices $\mathsf{C}^{1} : \mathcal{X} \to \mathcal{Y}^{1}$ and $\mathsf{C}^{2} : \mathcal{X} \to \mathcal{Y}^{2}$, their \emph{external probabilistic choice with fixed probability $r$} is the channel $(\mathsf{C}^1 \ \boxplus_{r}\ \mathsf{C}^2) : \mathcal{X} \to (\mathcal{Y}^{1} \uplus \mathcal{Y}^{2})$ defined as \cite[Def. 8.2]{Alvim2020}:
\begin{equation}
    \label{eq:external-choice}
    (\mathsf{C}^1 \ \boxplus_{r}\ \mathsf{C}^2)_{x,y} =
    \begin{cases}
        (1-r) \mathsf{C}_{x,y}^{1} & \quad \text{if } y \text{ from } \mathcal{Y}^{1} \text{,} \\
        r \mathsf{C}_{x,y}^{2} & \quad \text{if } y \text{ from } \mathcal{Y}^{2} \text{.}
    \end{cases}
\end{equation}

\paragraph{Theorem \ref{theo:kronecker-capacity}}
For channel matrices $\mathsf{C} : \mathcal{X} \to \mathcal{Y}$ and $\mathsf{D} : \mathcal{X'} \to \mathcal{Y'}$, the multiplicative (average-case) Bayes capacity of the Kronecker product $\mathsf{C} \otimes \mathsf{D} : (\mathcal{X}, \mathcal{X'}) \to (\mathcal{Y}, \mathcal{Y'})$ is:
\begin{equation*}
    \mathcal{ML}_{1}^{\times} (\vartheta'', \mathsf{C} \otimes \mathsf{D}) = \mathcal{ML}_{1}^{\times} (\vartheta, \mathsf{C}) \cdot \mathcal{ML}_{1}^{\times} (\vartheta', \mathsf{D}),
\end{equation*}
i.e. the product of the multiplicative Bayes capacities of $\mathsf{C}$ and $\mathsf{D}$.

\begin{proof}
    \begin{align*}
        &\mathcal{ML}_{1}^{\times} (\vartheta, \mathsf{C}) \cdot \mathcal{ML}_{1}^{\times} (\vartheta', \mathsf{D}) =\\
        &= \sum_{y \in \mathcal{Y}} \max_{x \in \mathcal{X}} \mathsf{C}_{x,y} \cdot \sum_{y' \in \mathcal{Y'}} \max_{x' \in \mathcal{X'}} \mathsf{D}_{x',y'} &\text{\qm{$\mathsf{C}, \mathsf{D}$ (cf.~\eqref{eq:bayes-capacity}).}}\\
        &= \sum_{y \in \mathcal{Y}} \sum_{y' \in \mathcal{Y'}} \max_{x \in \mathcal{X}} \mathsf{C}_{x,y} \cdot \max_{x' \in \mathcal{X'}} \mathsf{D}_{x',y'} &\text{\qm{Distributively.}}\\
        &= \sum_{\substack{y \in \mathcal{Y},\\y' \in \mathcal{Y'}}} \max_{\substack{x \in \mathcal{X},\\x' \in \mathcal{X'}}} \mathsf{C}_{x,y} \cdot \mathsf{D}_{x',y'} &\text{\qm{Independence of $x, w$.}}\\
        &= \sum_{\substack{y \in \mathcal{Y},\\y' \in \mathcal{Y'}}} \max_{\substack{x \in \mathcal{X},\\x' \in \mathcal{X'}}} (\mathsf{C} \otimes \mathsf{D})_{(x,x'),(y,y')} &\text{\qm{(cf.~\eqref{eq:kronecker-product}).}}\\
        &= \mathcal{ML}_{1}^{\times} (\vartheta'', \mathsf{C} \otimes \mathsf{D}) &\text{\qm{(cf.~\eqref{eq:bayes-capacity}).}}
    \end{align*}
\end{proof}

\paragraph{Corollary \ref{cor:kronecker-vulnerability}}
The Bayes vulnerability of the Kronecker product of channel matrices $\mathsf{C} : \mathcal{X} \to \mathcal{Y}$ and $\mathsf{D} : \mathcal{X'} \to \mathcal{Y'}$ is:
\begin{equation*}
    V_1 [\vartheta'' \triangleright (\mathsf{C} \otimes \mathsf{D})] = V_1 [\vartheta \triangleright \mathsf{C}] \cdot V_1 [\vartheta' \triangleright \mathsf{D}],
\end{equation*}
i.e. the product of the Bayes vulnerabilities of $\mathsf{C}$ and $\mathsf{D}$.

\begin{proof}
    Assume $\mathsf{C}$ is $m \times n$ and $\mathsf{D}$ is $p \times q$, then by Def.~\ref{def:kronecker-product}, $\mathsf{C} \otimes \mathsf{D}$ is $pm \times qn$.
    Hence,
    \begin{align*}
        &\mathcal{ML}_{1}^{\times} (\vartheta, \mathsf{C}) \cdot \mathcal{ML}_{1}^{\times} (\vartheta', \mathsf{D}) =\\
        &= \frac{V_1 [\vartheta \triangleright \mathsf{C}]}{V_1 (\vartheta)} \cdot \frac{V_1 [\vartheta' \triangleright \mathsf{D}]}{V_1 (\vartheta')} &\text{\qm{(cf.~\ref{eq:leakage}).}}\\
        &= m \cdot p \cdot V_1 [\vartheta \triangleright \mathsf{C}] \cdot V_1 [\vartheta' \triangleright \mathsf{D}] &\text{\qm{Assumption.}}\\
        &= \mathcal{ML}_{1}^{\times} (\vartheta'', \mathsf{C} \otimes \mathsf{D}) &\text{\qm{Thm.~\ref{theo:kronecker-capacity}.}}\\
        &= p \cdot m \cdot V_1[\vartheta'' \triangleright (\mathsf{C} \otimes \mathsf{D})] &\text{\qm{(cf.~\ref{eq:leakage}).}}\\
        &\therefore V_1[\vartheta'' \triangleright (\mathsf{C} \otimes \mathsf{D})] = V_1 [\vartheta \triangleright \mathsf{C}] \cdot V_1 [\vartheta' \triangleright \mathsf{D}] &\text{\qm{Above.}}\\
    \end{align*}
\end{proof}

\paragraph{Theorem \ref{theo:iba-gain-initial}}
The analyst prior expected gain considering the IBA gain function, $g_{\text{IBA}}$, and a prior probability distribution on top-$s$ sets, $\pi$, is given by the \emph{prior IBA vulnerability} (cf.~\eqref{eq:prior-vulnerability}):
\begin{equation*}
    V_{g_{\text{IBA}}} (\pi) = \max_{t \in \mathtt{T}} \sum_{\sigma \in \Sigma \ : \ t \in \sigma} \pi_{\sigma}.
\end{equation*}

\begin{proof}
    \begin{align*}
        &V_{g_{\text{IBA}}} (\pi) =\\
        &= \max_{t \in \mathtt{T}} \sum_{\sigma \in \Sigma} \pi_{\sigma} g_{\text{IBA}}(t,\sigma) &\text{\qm{(cf.~\ref{eq:prior-vulnerability}).}}\\
        &= \max_{t \in \mathtt{T}} \sum_{\sigma \in \Sigma \ : \ t \in \sigma} \pi_{\sigma} &\text{\qm{(cf.~\ref{eq:iba-gain}).}}
    \end{align*}
\end{proof}

\paragraph{Theorem \ref{theo:iba-gain-final-bounded-noise}}
The analyst posterior expected gain considering the IBA gain function, $g_{\text{IBA}}$, the bounded noise channel, $\mathsf{C}_{\mathit{BN}}$, and a prior probability distribution on top-$s$ sets, $\pi$, is given by the \emph{posterior IBA vulnerability for bounded noise} (cf.~\eqref{eq:posterior-vulnerability}):
\begin{equation*}
    V_{g_{\text{IBA}}} [\pi \triangleright \mathsf{C}_{\mathit{BN}}] = 1.
\end{equation*}

\begin{proof}
    \begin{align*}
        &V_{g_{\text{IBA}}} [\pi \triangleright \mathsf{C}_{\mathit{BN}}] =\\
        &= \sum_{t \in \mathtt{T}} \max_{t \in \mathtt{T}} \sum_{\sigma \in \Sigma} \pi_{\sigma} \mathsf{C}_{{\mathit{BN}}_{\sigma,t}} g_{\text{IBA}}(t,\sigma) &\text{\qm{(cf.~\ref{eq:posterior-vulnerability}).}}\\
        &= \sum_{t \in \mathtt{T}} \max_{t \in \mathtt{T}} \sum_{\sigma \in \Sigma \ : \ t \in \sigma} \pi_{\sigma} \mathsf{C}_{{\mathit{BN}}_{\sigma,t}} &\text{\qm{(cf.~\ref{eq:iba-gain}).}}\\
        &= \sum_{t \in \mathtt{T}} \max_{t \in \mathtt{T}} \sum_{\sigma \in \Sigma} \pi_{\sigma} \mathsf{C}_{{\mathit{BN}}_{\sigma,t}} &\text{\qm{$\mathsf{C}_{{\mathit{BN}}_{\sigma,t}} = 0$ if $t \notin \sigma$.}}\\
        &= \sum_{t \in \mathtt{T}} \sum_{\sigma \in \Sigma} \pi_{\sigma} \mathsf{C}_{{\mathit{BN}}_{\sigma,t}} &\text{\qm{$\sum_{t \in \mathtt{T}}$, $\max_{t \in \mathtt{T}}$, both over $t$.}}\\
        &= \sum_{\sigma \in \Sigma} \sum_{t \in \mathtt{T}} \pi_{\sigma} \mathsf{C}_{{\mathit{BN}}_{\sigma,t}} &\dagger\\
        &= \sum_{\sigma \in \Sigma} \pi_{\sigma} \sum_{t \in \mathtt{T}} \mathsf{C}_{{\mathit{BN}}_{\sigma,t}} &\text{\qm{Distributively.}}\\
        &= 1
    \end{align*}
    
    Step $\dagger$ is due to commutativity and associativity.
    
    Note that $\mathsf{C}_{\mathit{BN}}$ reports \emph{only} one of the genuine topics of interest for a user.
    This means that the adversary can be confident that the result, whatever it is, will yield $1$ with respect to the $g_{\text{IBA}}$ gain function.
\end{proof}

\paragraph{Lemma \ref{lem:internal-choice-lower-bound}}
Given compatible channel matrices $\mathsf{C}$ and $\mathsf{D}$, i.e. both with the same input set $\mathcal{X}$, and their internal fixed-probability choice with probability $r$, $\mathsf{C} \ \oplus_{r} \mathsf{D}$, the posterior vulnerability of their internal probabilistic choice is bounded below by the maximum posterior vulnerability of the individual channels:
\begin{equation*}
    V_{g} [\pi \triangleright \mathsf{C} \ \oplus_{r} \mathsf{D}] \geq \max \{ (1-r) \cdot V_{g} [\pi \triangleright \mathsf{C}], r \cdot V_{g} [\pi \triangleright \mathsf{D}] \}.
\end{equation*}

\begin{proof}
    We can write $(\mathsf{C} \ \oplus_{r} \mathsf{D})_{(x,y)} = (1-r) \cdot \mathsf{C}_{(x,y)} + r \cdot \mathsf{D}_{(x,y)}$, since if $\mathcal{Y}^{1} \neq \mathcal{Y}^{2}$ for channels $\mathsf{C} : \mathcal{X} \to \mathcal{Y}^{1}$ and $\mathsf{D} : \mathcal{X} \to \mathcal{Y}^{2}$ one can make $\mathcal{Y}^{1} = \mathcal{Y}^{2}$ by introducing into each channel their missing columns filled with zeroes.
    Hence,
    \begin{align*}
        &V_{g} [\pi \triangleright \mathsf{C} \ \oplus_{r} \mathsf{D}] =\\
        &= \sum_{y \in \mathcal{Y}} \max_{w \in \mathcal{W}} \sum_{x \in \mathcal{X}} \pi_{x} ((1-r) \mathsf{C}_{(x,y)} + r \mathsf{D}_{(x,y)}) g(w,x) &\text{\qm{(cf.~\ref{eq:posterior-vulnerability}).}}\\
        &= \sum_{y \in \mathcal{Y}} \max_{w \in \mathcal{W}} \left( \sum_{x \in \mathcal{X}} \pi_{x} (1-r) \mathsf{C}_{(x,y)} g(w,x) \right.\\
        &\left. \qquad\qquad\qquad + \sum_{x \in \mathcal{X}} \pi_{x} r \mathsf{D}_{(x,y)} g(w,x) \right) &\dagger\\
        &\geq \max \left\{ \sum_{y \in \mathcal{Y}} \max_{w \in \mathcal{W}} \sum_{x \in \mathcal{X}} \pi_{x} (1-r) \mathsf{C}_{(x,y)} g(w,x), \right.\\
        &\left. \qquad\qquad \sum_{y \in \mathcal{Y}} \max_{w \in \mathcal{W}} \sum_{x \in \mathcal{X}} \pi_{x} r \mathsf{D}_{(x,y)} g(w,x) \right\} &\ddagger\\
        &= \max \{ (1-r) \cdot V_{g} [\pi \triangleright \mathsf{C}], r \cdot V_{g} [\pi \triangleright \mathsf{D}] \} &\text{\qm{(cf.~\ref{eq:posterior-vulnerability}).}}
    \end{align*}
    
    Step $\dagger$ is due to summation commutativity and associativity.
    
    Step $\ddagger$ is due to both arguments of the $\max$ in $\dagger$ being greater than or equal to $0$.
\end{proof}

\paragraph{Theorem \ref{theo:iba-gain-final}}
The analyst posterior expected gain considering the IBA gain function, the final channel for utility analyses, $\mathsf{C}_{\mathit{BN} \oplus_{r} \mathit{DP}}$, and a prior probability distribution on top-$s$ sets, $\pi$, is given by the \emph{posterior IBA vulnerability for the Topics API} (cf.~\eqref{eq:posterior-vulnerability}):
\begin{equation*}
    (1-r) \leq V_{g_{\text{IBA}}} [\pi \triangleright \mathsf{C}_{\mathit{BN} \oplus_{r} \mathit{DP}}] \leq (1-r) + r \cdot V_{g_{\text{IBA}}} (\pi).
\end{equation*}

\begin{proof}
    First, for $(1-r) \leq V_{g_{\text{IBA}}} [\pi \triangleright \mathsf{C}_{\mathit{BN} \oplus_{r} \mathit{DP}}]$:
    \begin{align*}
        &V_{g_{\text{IBA}}} [\pi \triangleright \mathsf{C}_{\mathit{BN} \oplus_{r} \mathit{DP}}] \geq\\
        &\geq \max \{ (1-r) \cdot V_{g_{\text{IBA}}} [\pi \triangleright \mathsf{C}_{\mathit{BN}}], r \cdot V_{g_{\text{IBA}}} [\pi \triangleright \mathsf{D}_{\mathit{DP}}] \} &\text{\qm{Lem.~\ref{lem:internal-choice-lower-bound}.}}\\
        &= \max \{ (1-r) \cdot V_{g_{\text{IBA}}} [\pi \triangleright \mathsf{C}_{\mathit{BN}}], r \cdot V_{g_{\text{IBA}}} (\pi) \} &\dagger\\
        &= \max \{ (1-r), r \cdot V_{g_{\text{IBA}}} (\pi) \} &\text{\qm{Thm.~\ref{theo:iba-gain-final-bounded-noise}.}}\\
        &= (1-r) &\ddagger
    \end{align*}
    
    Step $\dagger$ is due to $V_{g_{\text{IBA}}} [\pi \triangleright \mathsf{C}_{\mathit{DP}}] = V_{g_{\text{IBA}}} (\pi)$, since the channel $\mathsf{C}_{\mathit{DP}}$ reveals nothing.
    
    Step $\ddagger$ is due to $r < 0.5$ and $V_{g_{\text{IBA}}} (\pi) \leq 1$.
    
    Second, for $V_{g_{\text{IBA}}} [\pi \triangleright \mathsf{C}_{\mathit{BN} \oplus_{r} \mathit{DP}}] \leq (1-r) + r \cdot V_{g_{\text{IBA}}} (\pi)$, we use the definitions of internal and external fixed-probability choices and of channel refinement \cite[Chapter 9]{Alvim2020}.
    
    We know that the external fixed-probability choice refines the internal fixed-probability choice, i.e. $V_{g} [\pi \triangleright \mathsf{C} \boxplus_{r} \mathsf{D}] \geq V_{g} [\pi \triangleright \mathsf{C} \oplus_{r} \mathsf{D}]$.
    Moreover, we can decompose the external fixed-probability choice as follows:
    \begin{align*}
        &V_{g_{\text{IBA}}} [\pi \triangleright \mathsf{C}_{\mathit{BN} \oplus_{r} \mathit{DP}}] \leq V_{g_{\text{IBA}}} [\pi \triangleright \mathsf{C}_{\mathit{BN} \boxplus_{r} \mathit{DP}}] =\\
        &= (1-r) \cdot V_{g_{\text{IBA}}} [\pi \triangleright \mathsf{C}_{\mathit{BN}}] + r \cdot V_{g_{\text{IBA}}} [\pi \triangleright \mathsf{C}_{\mathit{DP}}] &\text{\qm{(cf.~\eqref{eq:external-choice}).}}\\
        &= (1-r) \cdot V_{g_{\text{IBA}}} [\pi \triangleright \mathsf{C}_{\mathit{BN}}] + r \cdot V_{g_{\text{IBA}}} (\pi) &\dagger\\
        &= (1-r) + r \cdot V_{g_{\text{IBA}}} (\pi) &\text{\qm{Thm.~\ref{theo:iba-gain-final-bounded-noise}.}}
    \end{align*}
    
    Step $\dagger$ is due to $V_{g_{\text{IBA}}} [\pi \triangleright \mathsf{C}_{\mathit{DP}}] = V_{g_{\text{IBA}}} (\pi)$, since the channel $\mathsf{C}_{\mathit{DP}}$ reveals nothing.
    
    Notice that when the prior is uniform, the choice of how to resolve the nondeterminism in $V_{g_{\text{IBA}}}$ does not affect expected gain.
    This means that there is no disadvantage for this adversary to not know whether $\mathsf{C}_{\mathit{DP}}$ or $\mathsf{C}_{\mathit{BN}}$ is being used.
    Therefore, we have $V_{g_{\text{IBA}}} [\pi \triangleright \mathsf{C}_{\mathit{BN} \oplus_{r} \mathit{DP}}] = V_{g_{\text{IBA}}} [\pi \triangleright \mathsf{C}_{\mathit{BN} \boxplus_{r} \mathit{DP}}]$, and the upper bound is achieved for the uniform prior.
\end{proof}

\newpage

\section{Datasets}
\label{app:datasets}

\begin{table*}[t]
    \centering
    \begin{tabular}{|c|c|c|c|c|c|c|c|c|c|c|c|c|c|c|}
        \cline{3-15}
        \multicolumn{2}{c}{} & \multicolumn{4}{|c|}{Unique} & \multicolumn{9}{|c|}{Browsing history size} \\
        \hline
        \multicolumn{1}{|p{5.1em}|}{\centering AOL Dataset} & Rows & Users & URLs & Domains & Topics & Min. & 25\% & 50\% & 75\% & 95\% & 99\% & Max. & Mean & $\sigma$ \\
        \hline
        \hline
        \centering Original & 36\,389\,567 & 657\,426 & 1\,632\,789 & --- & --- & 1 & 5 & 17 & 52 & 228 & 566 & 279\,430 & 55.35 & 367.22 \\
        \hline
        \centering Treated & 19\,426\,293 & 521\,607 & --- & 1\,300\,484 & --- & 1 & 3 & 10 & 33 & 160 & 404 & 150\,802 & 37.24 & 226.67 \\
        \hline
        \multicolumn{1}{|p{5.1em}|}{\centering Reduced \\ (Citizen Lab)} & 3\,135\,270 & 342\,003 & --- & 4\,969 & 31 & 1 & 1 & 3 & 8 & 34 & 92 & 23\,505 & 9.17 & 46.00 \\
        \hline
        \multicolumn{1}{|p{5.1em}|}{\centering Reduced \\ (Google v1)} & 2\,493\,895 & 325\,383 & --- & 2\,718 & 171 & 1 & 1 & 3 & 7 & 28 & 71 & 19\,011 & 7.66 & 37.29 \\
        \hline
    \end{tabular}
    \caption{Treated and Reduced datasets statistics. The Original dataset is for reference only. Both datasets with topics classifications were derived from the Treated dataset. All datasets span the period from 2006/03/01 to 2006/05/31, inclusive.}
    \label{tab:datasets-treated}
\end{table*}

We briefly introduced the \emph{Experimental} AOL datasets in Sec.~\ref{sec:experimental-datasets}.
In this section, we further detail our data treatments and present the statistics for the \emph{Treated} and \emph{Reduced} AOL datasets.

The AOL search logs dataset from $2006$ \cite{Pass2006} comprises $36\,389\,567$ rows, one for each logged \texttt{Query} with the respective unique user identification number, \texttt{AnonID}, query date and time, \texttt{QueryTime}, visited URL, \texttt{ClickURL}, and its rank on the search results, \texttt{ItemRank}, accounting for $657\,426$ unique users and $1\,632\,789$ unique URLs.
The data spans the period from $2006/03/01$ to $2006/05/31$, inclusive.

As proposed by Roesner et al. \cite{Roesner2012}, this dataset may be used to reconstruct Internet users' browsing histories based on their visited URLs.
Browsing history size statistics for the \emph{Original} and \emph{Experimental} AOL datasets are presented in Tab.~\ref{tab:datasets-experiments} in Sec.~\ref{sec:experimental-datasets}, and for the \emph{Treated} and \emph{Reduced} AOL datasets are presented in Tab.~\ref{tab:datasets-treated}.

We have run our data treatments and experiments on a \texttt{Debian GNU/Linux trixie/sid} system running \texttt{Python v3.11.9} and the following Python packages: \texttt{bvmlib v1.0.0}, \texttt{numpy v1.24.3}, \texttt{pandas v2.0.1}, \texttt{qif v1.2.3}, \texttt{requests v2.31.0}, \texttt{tldextract v5.1.2}, and \texttt{urllib3 v1.26.16}.

\subsection{Treated AOL Dataset}
\label{app:datasets-treated}

We have started the data treatment from the Original AOL dataset by removing an ASCII control character (\texttt{\textbackslash x19}) found as a suffix to one of the \texttt{AnonID} values.
This was followed by the drop of unnecessary columns, i.e. \texttt{Query} and \texttt{ItemRank}, and of rows without a URL.
The Original AOL dataset includes $16\,946\,938$ rows with missing values on the \texttt{ClickURL} column.
This was expected, according to the Original AOL dataset documentation, when a search query was not followed by the user clicking on one of the results.
Also, a request for the next page of results would be logged as a new \texttt{Query} with an updated \texttt{QueryTime}.

We have then randomly remapped every \texttt{AnonID} value to a new \texttt{RandID} value in a non-retrievable way as to avoid direct linkage of individuals from our treated datasets to the Original AOL dataset.
This was achieved by using Python's \texttt{random.SystemRandom}\footnote{\url{https://docs.python.org/3/library/random.html\#random.SystemRandom}} class, which does not rely on software state and generates non-reproducible sequences.

Next, we have removed another two ASCII control characters (\texttt{\textbackslash x0e} and \texttt{\textbackslash x0f}) found as part of the \texttt{ClickURL} values of two distinct entries.
This was followed by the removal of two ASCII printable characters (\texttt{\textbackslash 5b} and \texttt{\textbackslash 5d}, corresponding to left and right square brackets, respectively), as they raise errors on Python's \texttt{urllib.parse.urlparse} function and are not expected as part of a domain name, which we want to derive from \texttt{ClickURL} values.

The following set of data treatments were performed to convert URLs to their respective domains and to fix inconsistencies.
We have first used Python's \texttt{urllib.parse.urlparse}\footnote{\url{https://docs.python.org/3/library/urllib.parse.html\#urllib.parse.urlparse}} function, which parses a URL into a named tuple with six components corresponding to \qm{the general structure of a URL},\footnote{A general URL: \qm{\texttt{scheme://netloc/path;parameters?query\#fragment}}.} to convert \texttt{ClickURL} values to either their \texttt{netloc} component or, in the case of an empty \texttt{netloc} component, to their \texttt{path} component.\footnote{According to Python's \texttt{urllib.parse.urlparse} documentation: \qm{Following the syntax specifications in RFC 1808 \cite{Fielding1995}, \texttt{urlparse} recognizes a \texttt{netloc} only if it is properly introduced by `//'. Otherwise the input is presumed to be a relative URL and thus to start with a \texttt{path} component.}}

We have then filtered the parsed results returned by Python's \texttt{urllib.parse.urlparse} function according to the definition of \emph{effective top-level domain} (eTLD),\footnote{\url{https://developer.mozilla.org/en-US/docs/Glossary/eTLD}} i.e. \qm{a domain under which domains can be registered by a single organization}, and to the technical requirements for domain names, i.e. \qm{a domain name consists of one or more labels, each of which is formed from the set of ASCII letters, digits, and hyphens (a–z, A–Z, 0–9, -), but not starting or ending with a hyphen} \cite{WikipediaContributors2004,Berners-Lee1994,Fielding1995}.

This was achieved by using Python's \texttt{tldextract}\footnote{\url{https://github.com/john-kurkowski/tldextract}} package to separate subdomain, domain, and public suffix, according to Mozilla's Public Suffix List,\footnote{\url{https://publicsuffix.org}} i.e. file \texttt{public\_suffix\_list.dat} from the repository \href{https://github.com/publicsuffix/list/tree/5e6ac3a082505ac4cf08858bdb38382d9a912833}{https://github.com/publicsuffix/list}, commit \texttt{5e6ac3a}, extended by the following discontinued \emph{top-level domains} (TLDs): \texttt{.bg.ac.yu}, \texttt{.ac.yu}, \texttt{.cg.yu}, \texttt{.co.yu}, \texttt{.edu.yu}, \texttt{.gov.yu}, \texttt{.net.yu}, \texttt{.org.yu}, \texttt{.yu}, \texttt{.or.tp}, \texttt{.tp}, and \texttt{.an}.

In the process, we have also checked the results for the following inconsistencies: if empty suffix, then dropped; if valid suffix with only one label but no domain, e.g. \qm{br} or \qm{com}, then dropped; if valid suffix with more than one label but no domain, e.g. \qm{gov.br}, then accepted; if valid suffix with more than one label and only \qm{w}s in the domain, e.g. \qm{www.gov.br}, then accepted suffix-only, i.e. \qm{gov.br}; if domain with invalid characters, e.g. \qm{+.gov.br} or \qm{-.gov.br}, then dropped; if valid suffix with one label and only \qm{w}s in the domain, e.g. \qm{www.br}, then dropped.

The Treated AOL dataset, file \texttt{AOL-treated.csv} \cite{Nunes2024b}, comprises $19\,426\,293$ rows, one for each visited \texttt{Domain}, with the respective random user identification number, \texttt{RandID}, and query date and time, \texttt{QueryTime}, accounting for $521\,607$ unique users and $1\,300\,484$ unique domains.
Browsing history size statistics are presented in Tab.~\ref{tab:datasets-treated}, and additional statistics, including number of \texttt{RandID}s per count of records and top \texttt{Domain} values by number of records, can be found in the file \texttt{AOL-data-treatment.ipynb} \cite{Nunes2024a}.\footnote{The unique domains from the Treated AOL dataset were also saved as an auxiliary dataset, file \texttt{AOL-treated-unique-domains.csv} \cite{Nunes2024b}.}

The following sets of data treatments were performed to assign topics to domains and to remove domains without a classification.
We have used two distinct static classifications: Citizen Lab's URL testing lists for Internet censorship \cite{CitizenLab2014}, presented in App.~\ref{app:datasets-reduced-citizen-lab}, and Google's static classification of domains provided with the Chrome browser, according to Google's Topics taxonomy v1, presented in App.~\ref{app:datasets-reduced-google}.
We are not evaluating machine classification models, so we have opted to use only human-based, static classifications.
We refer readers to \cite{Beugin2023} for empirical analyses of the sort.

\subsection{Reduced AOL Dataset with Citizen Lab's Classification}
\label{app:datasets-reduced-citizen-lab}

The original Citizen Lab URL testing lists dataset, i.e. the merge of all lists from the repository \href{https://github.com/citizenlab/test-lists/tree/ebd0ee8d41977b381972b2f6c471af5437d8d015/lists}{https://github.com/citizenlab/test-lists}, commit \texttt{ebd0ee8}, and using the new category codes, comprises $41\,209$ rows, one for each \texttt{url} with the respective \texttt{category\_code}, accounting for $33\,861$ unique URLs and $31$ distinct categories.
We have performed the same data treatment described in App.~\ref{app:datasets-treated} to convert URLs to their respective domains,\footnote{\texttt{Citizen-Lab-Classification-data-treatment.ipynb} \cite{Nunes2024a}.} resulting in $27\,792$ rows, one for each \texttt{domain} and the respective set of \texttt{topics}.\footnote{\texttt{Citizen-Lab-Classification.csv} \cite{Nunes2024b}.}

Matching the domains from the Citizen Lab classification with the unique domains from the Treated AOL dataset resulted in $4\,969$ unique domains, of which $3\,575$ are full matches and $1\,394$ are partial matches,\footnote{\texttt{AOL-treated-Citizen-Lab-Classification-domain-matching.ipynb} \cite{Nunes2024a}.} i.e. excluding the subdomain, e.g. \texttt{gps.gov.uk} in the Treated AOL dataset matched with \texttt{gov.uk} in the Citizen Lab URL testing list dataset.\footnote{\texttt{AOL-treated-Citizen-Lab-Classification-domain-match.csv} \cite{Nunes2024b}.}

The Reduced AOL dataset with Citizen Lab's classification, file \texttt{AOL-reduced-Citizen-Lab-Classification.csv} \cite{Nunes2024b}, was then derived from the Treated AOL dataset and comprises $3\,135\,270$ rows, one for each visited \texttt{Domain} with the respective \texttt{RandID}, \texttt{QueryTime}, and set of \texttt{topics}, accounting for $342\,003$ unique users and $4\,969$ unique domains.
Browsing history size statistics are presented in Tab.~\ref{tab:datasets-treated}, and additional statistics can be found in the file \texttt{AOL-reduced-Citizen-Lab-Classification.ipynb} \cite{Nunes2024a}.

\subsection{Reduced AOL Dataset with Google's Topics Classification}
\label{app:datasets-reduced-google}

The static classification of domains provided by Google with the Chrome browser, after extraction, comprises $9\,046$ rows, one for each \texttt{domain} and the respective set of \texttt{topics} by code, according to Google's Topics taxonomy v1, which accounts for $349$ distinct categories.\footnote{\url{https://github.com/patcg-individual-drafts/topics/blob/main/taxonomy_v1.md}}
We have treated this dataset only to remove domains without a classification,\footnote{\texttt{AOL-treated-Google-Topics-Classification-v1-domain-matching.ipynb} \cite{Nunes2024a}.} which matched with the unique domains from the Treated AOL dataset resulted in $2\,718$ unique domains,\footnote{\texttt{AOL-treated-Google-Topics-Classification-v1-domain-match.csv} \cite{Nunes2024b}.} of which $1\,444$ are full matches and $1\,274$ are partial matches.
Also, $101$ full matches without a classification were removed.

The Reduced AOL dataset with Google's Topics taxonomy v1, file \texttt{AOL-reduced-Google-Topics-Classification-v1.csv} \cite{Nunes2024b}, was then derived from the Treated AOL dataset and comprises $2\,493\,895$ rows, one for each visited \texttt{Domain} with the respective \texttt{RandID}, \texttt{QueryTime}, and set of \texttt{topics}, accounting for $325\,383$ unique users and $2\,718$ unique domains.
Browsing history size statistics are presented in Tab.~\ref{tab:datasets-treated}, and additional statistics are in the file \texttt{AOL-reduced-Google-Topics-Classification-v1.ipynb} \cite{Nunes2024a}.

\subsection{Set of Experimental AOL Datasets}
\label{app:datasets-experimental-set}

The experimental results reported in Sec.~\ref{sec:experimental} were obtained using the set of Experimental AOL datasets, which were derived from the Treated and Reduced AOL datasets and received additional data treatments for consistency with the modeled scenarios and for better reproducibility of results.

First, we have dropped all users who have visited only one domain, including users who have only visited the same domain multiple times.
Such users (singletons), by definition, would not be vulnerable to cross-site tracking, neither by third-party cookies, nor by the Topics API.

Second, we have dropped one outlier user who accounted for as many as $150\,802$ domain visits in an interval of three months.
This accounts for $69$ domain visits per hour, $24$ hours a day, for $91$ days.
Meanwhile, all other users have at most $6\,227$ domain visits in the same interval of time.

Third, we have made the (data-dependent) construction of the top-$s$ sets of topics for each user as part of the dataset generation instead of the analyses.
This does not simulate the data-dependent aspects of the Topics API but allows for exactly reproducible results and to empirically verify code correctness.\footnote{\texttt{QIF-analyses-AOL-experimental.ipynb} \cite{Nunes2024a}.}\footnote{\texttt{QIF-analyses-AOL-experimental-Citizen-Lab.ipynb} \cite{Nunes2024a}.}\footnote{\texttt{QIF-analyses-AOL-experimental-Google-Topics-v1.ipynb} \cite{Nunes2024a}.}
Note that we also provide the code that simulates the data-dependent aspects of the Topics API and that take as input the Treated and Reduced AOL datasets instead of the Experimental AOL datasets.\footnote{\texttt{QIF-analyses-AOL-treated.ipynb} \cite{Nunes2024a}.}\footnote{\texttt{QIF-analyses-AOL-reduced-Citizen-Lab.ipynb} \cite{Nunes2024a}.}\footnote{\texttt{QIF-analyses-AOL-reduced-Google-Topics-v1.ipynb} \cite{Nunes2024a}.}

Finally, we have merged users' records by \texttt{RandID} to define their browsing histories as lists of tuples, with each tuple containing a visited domain and the respective date and time, and their lists of all topics and of top-$s$ topics, when applicable.

\subsubsection{Experimental AOL Dataset}
\label{app:datasets-experimental}

The removal of singletons from the Treated AOL dataset accounted for $85\,601$ users and their corresponding $120\,512$ records, i.e. individual domain visits, while the removal of the outlier user accounted for $150\,802$ records.

The Experimental AOL dataset, file \texttt{AOL-experimental.csv} \cite{Nunes2024b}, comprises $436\,005$ rows, one for each user (\texttt{RandID}), with the respective \texttt{BrowsingHistory} list.
This accounts for $19\,154\,979$ records, i.e. individual domain visits, and for $1\,291\,534$ unique domains.
Browsing history size statistics are presented in Tab.~\ref{tab:datasets-experiments}, and additional statistics, including number of \texttt{RandID}s per count of records and top \texttt{Domain} values by number of records, can be found in the file \texttt{AOL-experimental.ipynb} \cite{Nunes2024a}.

\subsubsection{Experimental AOL Dataset with Citizen Lab's Classification}
\label{app:datasets-experimental-citizen-lab}

The removal of singletons from the Reduced AOL dataset with Citizen Lab's classification accounted for $130\,689$ users and their corresponding $264\,187$ records, i.e. individual domain visits, while the removal of the outlier user accounted for $23\,505$ records.

The Experimental AOL dataset with Citizen Lab's classification, file \texttt{AOL-experimental-Citizen-Lab-Classification.csv} \cite{Nunes2024b}, comprises $211\,313$ rows, one for each user (\texttt{RandID}), with the respective \texttt{BrowsingHistory}, \texttt{AllTopics}, and \texttt{sTopics} lists.
This accounts for $2\,847\,578$ records, i.e. individual domain visits, and for $4\,872$ unique domains.
Browsing history size statistics are presented in Tab.~\ref{tab:datasets-experiments}, and additional statistics can be found in the file \texttt{AOL-experimental-Citizen-Lab-Classification.ipynb} \cite{Nunes2024a}.

\subsubsection{Experimental AOL Dataset with Google's Topics Classification}
\label{app:datasets-experimental-google}

The removal of singletons from the Reduced AOL dataset with Google's Topics classification accounted for $127\,359$ users and their corresponding $231\,224$ records, i.e. individual domain visits, while the removal of the outlier user accounted for $19\,011$ records.

\sloppy

The Experimental AOL dataset Google's Topics classification, file \texttt{AOL-experimental-Google-Topics-Classification-v1.csv} \cite{Nunes2024b}, comprises $198\,023$ rows, one for each user (\texttt{RandID}), with the respective \texttt{BrowsingHistory}, \texttt{AllTopics}, and \texttt{sTopics} lists.
This accounts for $2\,243\,660$ records, i.e. individual domain visits, and for $2\,652$ unique domains.
Browsing history size statistics are presented in Tab.~\ref{tab:datasets-experiments}, and additional statistics can be found in the file \texttt{AOL-experimental-Google-Topics-Classification-v1.ipynb} \cite{Nunes2024a}.

\fussy

\section{Longitudinal Model}
\label{app:model-longitudinal}

\begin{figure*}[t]
    \centering
    \includegraphics[width=0.95\textwidth]{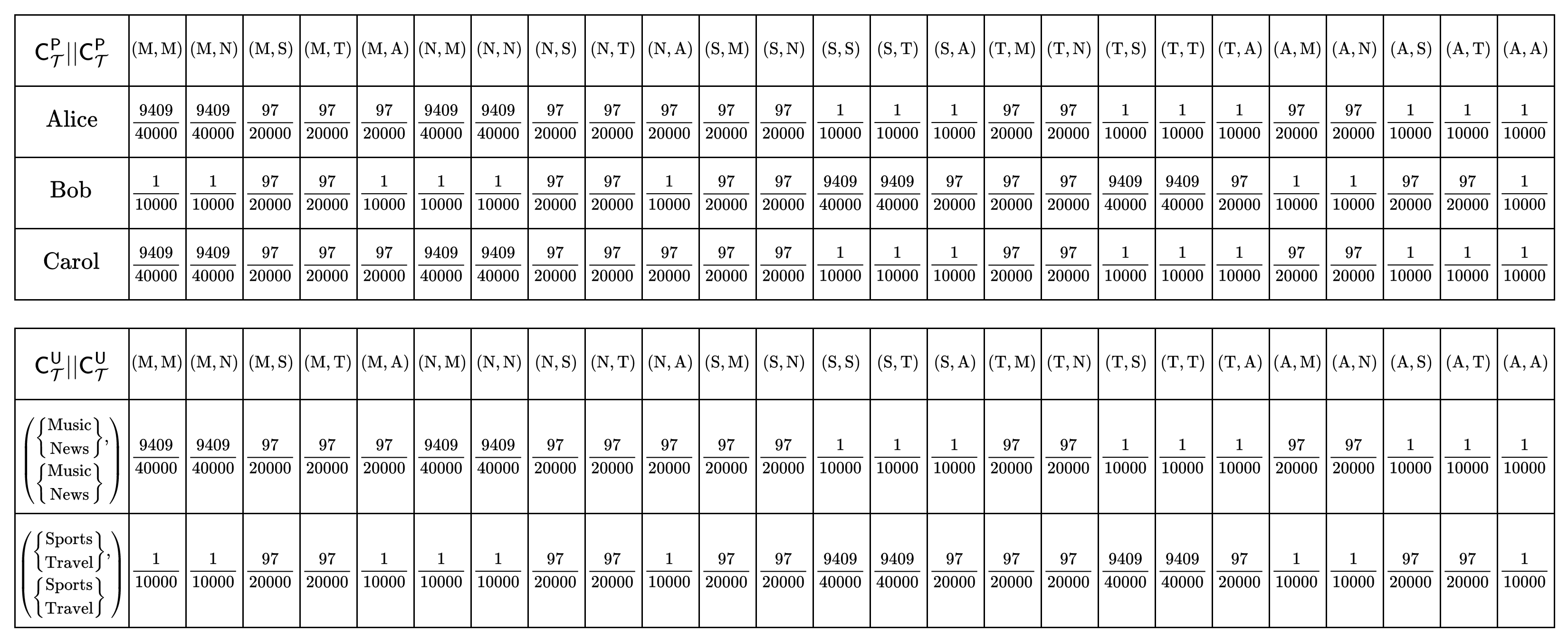}
    \caption{Final channels for privacy and utility analyses for two epochs. The channel $\mathsf{C}_{\mathcal{T}}^{\mathsf{P}}$ is the cascading of channels $\mathsf{C}_{\mathit{BH}} \mathsf{C}_{G} \mathsf{C}_{\mathit{BN} \oplus_{0.05} \mathit{DP}}$, as defined in Fig.~\ref{fig:privacy-channels}, and the channel $\mathsf{C}_{\mathcal{T}}^{\mathsf{U}} = \mathsf{C}_{\mathit{BN} \oplus_{0.05} \mathit{DP}}$, as defined in Fig.~\ref{fig:topics-channels}. The binary operator $\cdot\ ||\ \cdot$ represents the parallel composition of two channels (cf.~\ref{eq:parallel-composition}). The composition $\mathsf{C}_{\mathcal{T}}^{\mathsf{P}} || \mathsf{C}_{\mathcal{T}}^{\mathsf{P}}$ represents the final channel for privacy analyses and $\mathsf{C}_{\mathcal{T}}^{\mathsf{U}} || \mathsf{C}_{\mathcal{T}}^{\mathsf{U}}$ represents the final channel for utility analyses. We consider the parallel compositions of identical channels for simplicity. The letters $\text{M}$, $\text{N}$, $\text{S}$, $\text{T}$, and $\text{A}$ stand for the topics $\text{Music}$, $\text{News}$, $\text{Sports}$, $\text{Travel}$, and $\text{Ads}$, respectively.}
    \Description{A probabilistic channel matrix representing the parallel composition of two identical channels used for privacy analyses maps Internet users to tuples of two topics. Another probabilistic channel matrix representing the parallel composition of other two identical channels used for utility analyses maps tuples of two sets of two topics to tuples of two topics.}
    \label{fig:topics-channels-parallel}
\end{figure*}

We start by introducing an additional QIF definition.

\paragraph{Parallel Composition}
This composition of two compatible channel matrices, i.e. both with the same input set $\mathcal{X}$, models a scenario in which an adversary has access to the outputs of both channels, each operating independently on the same secret.
Formally, given channel matrices $\mathsf{C}^{1} : \mathcal{X} \to \mathcal{Y}^{1}$ and $\mathsf{C}^{2} : \mathcal{X} \to \mathcal{Y}^{2}$, their parallel composition $\mathsf{C}^{1} || \mathsf{C}^{2} : \mathcal{X} \to \mathcal{Y}^{1} \times \mathcal{Y}^{2}$ is defined as \cite[Def. 8.1]{Alvim2020}:
\begin{equation}
    \label{eq:parallel-composition}
    ( \mathsf{C}^{1} || \mathsf{C}^{2} )_{x, (y_{1},y_{2})} := \mathsf{C}_{x,y_{1}}^{1} \times \mathsf{C}_{x,y_{2}}^{2},
\end{equation}
for all $x \in \mathcal{X}$, $y_{1} \in \mathcal{Y}^{1}$, and $y_{2} \in \mathcal{Y}^{2}$.

We use this concept to model the longitudinal aspect of the Topics API when considering more than one epoch.
For instance, as a simplified example, we consider next the parallel composition of identical channels.

For the privacy analyses, we define channel $\mathsf{C}_{\mathcal{T}}^{\mathsf{P}} = \mathsf{C}_{\mathcal{T}}$, according to the channel cascading defined in Fig.~\ref{fig:privacy-channels} for channel $\mathsf{C}_{\mathcal{T}}$.
Hence, the parallel composition of $\mathsf{C}_{\mathcal{T}}^{\mathsf{P}}$ with itself (cf.~\ref{eq:parallel-composition}), i.e. $\mathsf{C}_{\mathcal{T}}^{\mathsf{P}} || \mathsf{C}_{\mathcal{T}}^{\mathsf{P}}$, is the channel presented at the top in Fig.~\ref{fig:topics-channels-parallel}.

For the utility analyses, we define channel $\mathsf{C}_{\mathcal{T}}^{\mathsf{U}} = \mathsf{C}_{\mathit{BN} \oplus_{0.05} \mathit{DP}}$, as in Fig.~\ref{fig:topics-channels}.
Hence, the parallel composition of $\mathsf{C}_{\mathcal{T}}^{\mathsf{U}}$ with itself (cf.~\ref{eq:parallel-composition}), i.e. $\mathsf{C}_{\mathcal{T}}^{\mathsf{U}} || \mathsf{C}_{\mathcal{T}}^{\mathsf{U}}$, is the channel presented at the bottom in Fig.~\ref{fig:topics-channels-parallel}.

Once the channels for the parallel compositions are defined, we can compute the same privacy and utility measures defined in Sec.~\ref{sec:theoretical}.
This is valid for the composition of as many channels as desired or as computationally treatable.

\end{document}